\documentclass{iopjournal}
\usepackage[backend=biber,style=ieee, maxnames=3]{biblatex}
\usepackage{siunitx}
\usepackage[nolist]{acronym}
\usepackage{booktabs} 
\DeclareUnicodeCharacter{2009}{\,}

\begin{document}
\newacro{ADC}{analog-to-digital converter}
\newacroplural{ADC}[ADCs]{analog-to-digital converters}

\newacro{DAC}{digital-to-analog converter}
\newacroplural{DAC}[DACs]{digital-to-analog converters}

\newacro{DRAM}{dynamic random-access memory}
\newacroplural{DRAM}[DRAMs]{dynamic random-access memories}

\newacro{IMC}{in-memory computing}
\newacroplural{IMC}[IMCs]{in-memory computing systems}

\newacro{GPU}{graphics processing unit}
\newacroplural{GPU}[GPUs]{graphics processing units}

\acrodef{LB}{Low Bound}
\acrodef{HB}{High Bound}
\acrodef{GM}{Memristor Conductance}

\acrodef{SALM}{Strong-Arm Latched Memristor}

\acrodef{MLSA}{Match-Line Sense Amplifier}
\acrodef{PCSA}{Precharge Sense Amplifier}

\acrodef{6T2M}{Six-Transistor Two-Memristor}

\newacro{HBM}{high-bandwidth memory}
\newacroplural{HBM}[HBMs]{high-bandwidth memories}

\newacro{KV}{key-value}
\newacroplural{KV}[KVs]{key-value pairs}

\newacro{LLM}{large language model}
\newacroplural{LLM}[LLMs]{large language models}

\newacro{MAC}{multiply-accumulate}
\newacroplural{MAC}[MACs]{multiply-accumulate operations}

\newacro{PWM}{pulse-width modulation}
\newacroplural{PWM}[PWMs]{pulse-width modulations}

\newacro{QAT}{quantization-aware training}
\newacroplural{QAT}[QATs]{quantization-aware training strategies}

\newacro{ReLU}{rectified linear unit}
\newacroplural{ReLU}[ReLUs]{rectified linear units}

\newacro{RNN}{recurrent neural network}
\newacroplural{RNN}[RNNs]{recurrent neural networks}

\newacro{SRAM}{static random-access memory}
\newacroplural{SRAM}[SRAMs]{static random-access memories}

\newacro{TPU}{tensor processing unit}
\newacroplural{TPU}[TPUs]{tensor processing units}

\newacro{VTC}{voltage-to-time converter}
\newacroplural{VTC}[VTCs]{voltage-to-time converters}

\newacro{WE}{write enable}
\newacroplural{WE}[WEs]{write enable signals}

\newacro{WL}{word line}
\newacroplural{WL}[WLs]{word lines}

\newacro{BL}{bit line}
\newacroplural{BL}[BLs]{bit lines}

\newacro{IP}{intellectual property}

\newacro{eFLASH}{embedded flash memory}

\newacro{CAM}{content-addressable memory}
\newacroplural{CAM}[CAMs]{content-addressable memories}

\newacro{aCAM}{analog content-addressable memory}
\newacroplural{aCAM}[aCAMs]{analog content-addressable memories}

\newacro{CIM}{compute-in-memory}
\newacroplural{CIM}[CIMs]{compute-in-memory architectures}

\newacro{ML}{match line}
\newacroplural{ML}[MLs]{match lines}

\newacro{DL}{data line}
\newacroplural{DL}[DLs]{data lines}

\newacro{PDK}{process design kit}

\newacro{SPICE}{simulation program with integrated circuit emphasis}

\newacro{CMOS}{complementary metal-oxide-semiconductor}

\newacro{MOSFET}{metal-oxide-semiconductor field-effect transistor}
\newacroplural{MOSFET}[MOSFETs]{metal-oxide-semiconductor field-effect transistors}

\newacro{HRS}{high-resistance state}
\newacro{LRS}{low-resistance state}

\newacro{SET}{set (programming voltage)}
\newacro{RESET}{reset (programming voltage)}

\newacro{IRIS}{Iris flower dataset}

\newacro{LUT}{lookup table}
\newacroplural{LUT}[LUTs]{lookup tables}

\newacro{3T1M}{three-transistor one-memristor}
\newacro{8T2M}{eight-transistor two-memristor}

\newacro{FSM}{finite state machine}
\newacroplural{FSM}[FSMs]{finite state machines}

\newacro{AI}{artificial intelligence}

\newacro{GAI}{generative artificial intelligence}

\newacro{SME}{small and medium enterprise}
\newacroplural{SME}{small and medium enterprises}

\newacro{VMM}{vector–matrix multiplication}
\newacroplural{VMM}[VMMs]{vector–matrix multiplications}

\newacro{SDK}{Software Development Kit}
\newacroplural{SDKs}{Software Development Kits}

\newacro{CPU}{central processing unit}

\newacro{NVM}{non-volatile memory}

\newacro{ReRAM}{resistive random-access memory}

\newacro{PCM}{phase-change memory}

\newacro{RISC-V}{Reduced Instruction Set Computer – Version V}
\newacroplural{RISC-V}[RISC-Vs]{Reduced Instruction Set Computer – Version V processors}

\newacro{PE}{Processing Element}
\newacroplural{PE}[PEs]{Processing Elements}

\newacro{NOC}{Network-on-Chip}
\newacroplural{NOC}[NOCs]{Networks-on-Chip}

\newacro{CIM}{Compute-in-Memory}
\newacroplural{CIM}[CIMs]{Compute-in-Memory systems}

\newacro{eFLASH}{embedded Flash memory}
\newacroplural{eFLASH}[eFLASH memories]{embedded Flash memories}

\newacro{DAC}{Digital-to-Analog Converter}
\newacroplural{DAC}[DACs]{Digital-to-Analog Converters}

\newacro{ADC}{Analog-to-Digital Converter}
\newacroplural{ADC}[ADCs]{Analog-to-Digital Converters}

\newacro{SRAM}{Static Random-Access Memory}
\newacroplural{SRAM}[SRAMs]{Static Random-Access Memories}

\newacro{MAC}{Multiply–Accumulate}
\newacroplural{MAC}[MACs]{Multiply–Accumulate operations}

\newacro{KV}{Key–Value}
\newacroplural{KV}[KV]{Key–Value pairs or caches}

\newacro{MRAM}{magnetoresistive random-access memory}

\newacro{MTJ}{magnetic tunnel junction}

\newacro{FeRAM}{ferroelectric random-access memory}

\newacro{FTJ}{ferroelectric tunnel junction}

\newacro{FeFET}{ferroelectric field-effect transistor}

\newacro{OSFET}{oxide semiconductor field-effect transistor}

\newacro{IGZO}{indium gallium zinc oxide}

\newacro{ITO}{indium tin oxide}

\newacro{NLP}{natural language processing}

\newacro{ANN}{artificial neural network}
\newacroplural{ANN}[ANNs]{artificial neural networks}

\newacro{BMFTR}{Bundesministerium für Forschung, Technologie und Raumfahrt}

\newacro{AE}{active electrode}

\newacro{OE}{ohmic electrode}

\articletype{Paper} 

\title{OTTER -- Two Transistor -- One RRAM Architecture for Reliable In-Memory-Computing in 28 nm CMOS Technology}

\author{Yang Chen$^{1,5}$\orcid{0009-0006-6511-1176}, 
Daniele Storelli$^1$\orcid{0009-0004-5623-7442},
Xinyi Zhao$^{1,5}$\orcid{0009-0000-1651-2922}, 
Ankit Bende$^{1,5}$\orcid{0009-0008-6434-7667}, 
Paul-Philipp Manea$^{3,5}$\orcid{0000-0001-6998-3066}, 
Oliver Artner$^{1,5}$\orcid{0009-0003-4041-5649}, 
Arun Ashok$^2$\orcid{0000-0001-5683-4352}, 
Kay Winterberg$^2$\orcid{0000-0003-2443-9627}, 
Godwin Paul$^{1,5}$\orcid{0009-0003-1238-4945},
Siyuan Jia$^5$\orcid{0009-0005-3468-9023},
Christian Roth$^2$\orcid{0000-0001-6669-9799}, 
Sabitha Kusuma$^2$\orcid{0000-0001-9801-8302}, 
Michael Schiek$^{3}$\orcid{0000-0003-0438-362X}, 
Vikas Rana$^{4}$\orcid{0000-0001-5432-0286}, 
Dirk Wouters$^5$\orcid{0000-0002-6766-8553},
Stephan Menzel$^1$\orcid{0000-0002-4258-2673}, 
André Zambanini$^2$\orcid{0000-0002-1585-4397}, 
Christian Grewing$^2$\orcid{0000-0003-1770-340X}, 
Stefan Wiefels$^{1,*}$\orcid{0000-0003-2820-9677}, 
Susanne Hoffmann-Eifert$^1$\orcid{0000-0003-1682-826X},
John Paul Strachan$^{3,5}$\orcid{0000-0002-1382-3677},
Stefan van Waasen$^{2}$\orcid{0000-0003-0682-7941}, 
and Regina Dittmann$^{1,4,5}$\orcid{0000-0003-1886-1864}}

\affil{$^1$Peter Grünberg Institute "Electronic Materials" (PGI-7), Forschungszentrum Jülich GmbH and JARA-FIT, Jülich, Germany}

\affil{$^2$Peter Grünberg Institute "Integrated Computing Architectures" (PGI-4), Forschungszentrum Jülich GmbH, Jülich, Germany}

\affil{$^3$Peter Grünberg Institute "Neuromorphic Compute Nodes" (PGI-14), Forschungszentrum Jülich GmbH, Jülich, Germany}

\affil{$^4$JARA Institute Energy-efficient information technology (PGI-10), Forschungszentrum Jülich GmbH, Jülich, Germany}

\affil{$^5$RWTH Aachen University, Faculty of Electrical Engineering and Information Technology, 52062 Aachen, Germany}

\affil{$^*$Author to whom any correspondence should be addressed.}

\email{s.wiefels@fz-juelich.de}


\begin{abstract}
This work presents \textit{OTTER}, a 28\,nm CMOS platform co-integrated with TaO$_x$-based valence-change mechanism (VCM) RRAM, demonstrating a two-transistor-one-memristive\allowbreak-device (2T1R) architecture for reliable in-memory computing. The 2T1R cell combines a low-drive-current (LD) transistor and a high-drive-current (HD) transistor in parallel, providing dedicated bias paths for SET programming and RESET operation, respectively. Through systematic experimental and simulated comparison of various transistor-pairing configurations using the physical compact model JART VCM Rth, design guidelines for transistor sizing are derived, establishing the minimum RESET transistor \textit{W}/\textit{L} required for complete RESET as a function of the SET current compliance. The 2T1R cell is further characterized under pulse-based programming, demonstrating multilevel analog conductance tuning with narrow, well separated conductance states across six programmable levels. An analog content-addressable memory (aCAM) design based on the same 2T1R cell is additionally analyzed at the circuit level, evaluating trade-offs between top- and bottom-connected RRAM comparator configurations. A hardware implementation of compute-in-memory (CIM) multiply-and-accumulate (MAC) operations is further demonstrated on a $15 \times 15$ 2T1R crossbar array.
\end{abstract}

\section{Introduction}

Neuromorphic and in-memory computing architectures have emerged as promising approaches to overcome the limitations of conventional von Neumann systems in data-intensive workloads such as machine learning and artificial intelligence~\cite{Mehonic2024002,zouBreakingNeumannBottleneck2021,burrNeuromorphicComputingUsing2017,ielminiResistiveSwitchingRandomAccess2025}. By performing computation directly within memory arrays, these architectures aim to reduce data transfer between processing and memory units, which significantly improves energy efficiency and computational throughput~\cite{ielminiInmemoryComputingResistive2018}. Among the various technologies investigated for realizing such systems, resistive random-access memory (RRAM) has attracted considerable attention due to its nano-scale footprint, high endurance and retention~\cite{nailUnderstandingRRAMEndurance2016,zahoorResistiveRandomAccess2020,wongMetalOxideRRAM2012a}, as well as its analog programmability~\cite{wanComputeinmemoryChipBased2022,ambrogioEquivalentaccuracyAcceleratedNeuralnetwork2018}, which enables the implementation of content-addressable memory (CAM) as well as compute-in-memory (CIM). While both CAM and CIM exploit the proximity of storage and computation, they target different functionalities. CAM enables fast, parallel search by directly matching input data against stored patterns~\cite{halawaniRRAMbasedCAMCombined2021,pedrettiInMemoryComputingResistive2021}, whereas CIM focuses on efficient execution of arithmetic operations, such as vector-matrix multiplication (VMM)~\cite{wanComputeinmemoryChipBased2022,legallo64coreMixedsignalInmemory2023}.\\
One of the most promising candidates among RRAM technologies is filamentary-type valence-change mechanism (VCM) based RRAM~\cite{Dittmann2022001,waserNanoionicsbasedResistiveSwitching2007}. It typically relies on the formation of a conductive filament in a dielectric switching layer, commonly based on metal oxides such as TaO\textsubscript{x}~\cite{qinTaOxBasedRRAMImproved2021,leeFastHighenduranceScalable2011}, HfO\textsubscript{x}~\cite{ielminiResistiveSwitchingRandomAccess2025,niuGeometricConductiveFilament2016,govoreanu10x10nm2HfHfOx2011} and TiO\textsubscript{x}~\cite{huInvestigationResistiveSwitching2022,shenAdvancesRRAMDevices2020}. During the forming process, defects such as oxygen vacancies or cation interstitials are generated  which act as mobile donors. By relocating these donors, the device conductance can be modulated in two directions: SET with increasing conductance towards a high-conductive state (HCS), also referred to as the low-resistance state (LRS), and RESET with decreasing conductance towards a low-conductive state (LCS), also referred to as the high-resistance state (HRS). The filament formation as well as the SET process are strongly current-dependent and therefore require precise current control. In addition, array structures require a selective element to suppress sneak currents during read and write operations~\cite{seokReviewThreeDimensionalResistive2014}. A widely adopted approach to address both requirements is the integration of the RRAM device with a transistor, forming the so-called one-transistor-one-memristive-device (1T1R) cell~\cite{chenEnduranceRetentionTradeoff2013,ielminiResistiveSwitchingRandomAccess2025}.
The 1T1R configuration has been extensively investigated in recent years and represents a standard building block for many RRAM-based memory and computing architectures~\cite{ResistiveMemoryBasedInMemory}.\\ Whereas the transistor is desired as a current limiter during forming and SET, it can hinder the RESET operation due to its series resistance. The resulting voltage divider has been shown to cause RESET failures if the voltage drop across the RRAM device becomes insufficient~\cite{Wiefels2023002, Kopperberg2022002}. In particular, if the RRAM device is to be tuned to multiple states over a broad range of conductance, this causes conflicting constraints on the transistor design~\cite{artnerInfluenceResidualIon2026}. Whereas a transistor with high drive current (HD) is beneficial to ensure sufficient voltage across the RRAM device during RESET, a transistor with low drive current (LD) enables finer conductance control during programming. To decouple these requirements, we explore a 2T1R structure in which an HD and an LD transistor are placed in parallel and connected in series with the RRAM device, each independently addressable via a dedicated word line.\\
We introduce the chip and research platform \textit{OTTER}\footnote{OTTER: Tw\textbf{\underline{o}} \textbf{\underline{T}}ransis\textbf{\underline{t}}or -- One M\textbf{\underline{e}}mristo\textbf{\underline{r}} Architecture for Reliable In-Memory-Computing at 28\,nm}, designed and fabricated in a \SI{28}{\nano\meter} CMOS technology and co-integrated with TaO\textsubscript{x}-based RRAM devices. The chip provides 2T1R test structures and arrays designed for CAM and CIM applications.\\
We present a systematic characterization of RRAM operation in the 2T1R configuration, covering DC sweep measurements, pulse-based multilevel programming, and array-level compute-in-memory operations. The DC characterization is supported by simulations using the physical compact model JART VCM Rth~\cite{sonComprehensiveCompactModel2025}, from which design guidelines for transistor sizing in 2T1R cells are derived.

\section{Design}
\label{sec:Design}

\begin{figure}[tbh]
    \centering
    \includegraphics[width=0.5\linewidth]{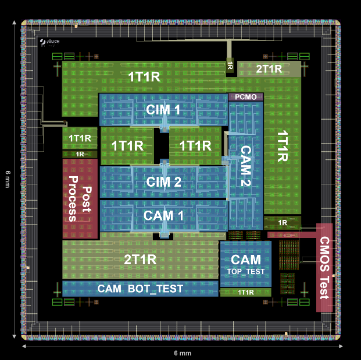}
    \caption{$6\times6\,\,\text{mm}^2$ \textit{OTTER} chip. Test structures for CMOS components and post processing are marked in red, blocks of single 1T1R and 2T1R structures are highlighted in green. Blue areas contain CIM and CAM arrays as well as respective test structures.}
    \label{fig:d0design}
\end{figure}

This section details the structures implemented on \textit{OTTER}, including 2T1R test structures as well as CAM and CIM arrays. \textit{OTTER} is implemented on dies of 6\,mm\,$\times$\,6\,mm in 28~nm CMOS technology as shown in Figure~\ref{fig:d0design}.

\subsection{2T1R structures}
The 2T1R architecture employs two dedicated access transistors, enabling the electrical requirements of SET and RESET operations to be decoupled. As shown in Figure~\ref{fig:CIMdesign}~a), the 2T1R cell places a high-drive-current (HD) transistor with a width-to-length ratio ($W/L$) of 10 and a low-drive-current (LD) transistor with a $W/L$ ratio of 2 in parallel, connected  in series with the RRAM device. Both transistors share the same bit line (BL) and source line (SL), while each transistor is independently addressed via its dedicated word line (WL). The HD transistor ensures a reliable RESET operation with minimized series resistance, whereas the LD transistor provides more precise current control during SET. Both are I/O transistors from TSMC 28\,nm technology. This 2T1R scheme provides independent control over the SET and RESET current paths within a single cell, offering additional design flexibility for balancing programming precision and RESET reliability. For reference, \textit{OTTER} also integrates 1T1R structures with either an HD or an LD transistor; their characterization is provided in the Supplementary Information. 
\subsection{CAM arrays}
\label{sec:cam_intro}
\begin{figure}[htb!]
    \centering
    \includegraphics[width=1\linewidth]{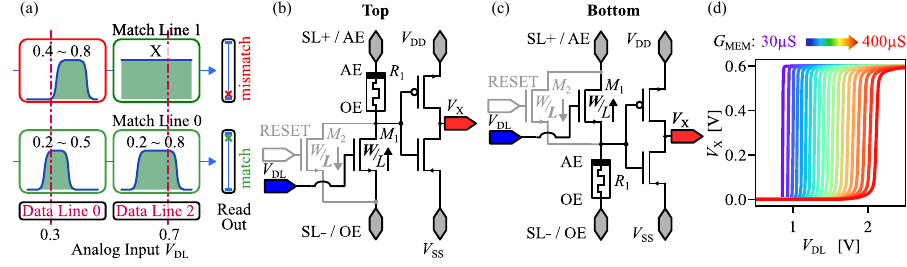}
    \caption{(a) Operation principle of an \ac{aCAM}. Each \ac{aCAM} cell stores a lower and upper bound defining a valid input interval (highlighted in green). If the applied input voltage lies within this interval, the cell returns a match; otherwise, a mismatch. All cells of a word share a common Match Line (ML) in a NOR configuration, such that a single mismatching cell causes a word-level mismatch. (b) Schematic of the memristive device comparator with the ohmic electrode (OE) connected to the access transistor $M_1$ (top-connected variant). (c) Schematic of the corresponding bottom-connected variant, in which the active electrode (AE) is connected to the access transistor $M_1$. (d) SPICE-simulated transfer characteristics of the memristive device comparator (Top orientation), showing the output voltage $V_\mathrm{X}$ as a function of the swept input voltage $V_\mathrm{DL}$ for memristive device conductances $G_\mathrm{MEM}$ ranging from 30\,\textmu S to 400\,\textmu S.}
    \label{fig:aCAM_intro}
\end{figure}

\Acp{CAM} are associative memories that perform parallel searches by comparing query data against all stored entries simultaneously. While conventional \acp{CAM} operate on binary patterns, \acp{aCAM} store analog ranges using multilevel memristive device conductance states, enabling each cell to determine whether an input lies within a programmable interval~\cite{Li2020}. As illustrated in Figure~\ref{fig:aCAM_intro}~a), the \ac{ML} combines all cell outputs, and a single mismatch causes the entire row to be marked as a mismatch. This enables continuous-valued similarity search and multi-interval matching, making \acp{aCAM} attractive for applications including Decision Trees and Random Forests~\cite{Pedretti2021tree,10753423}, Finite State Machines~\cite{Graves2022}, DNA Pattern Matching~\cite{he2024shiftcam}, Transformer Attention Mechanisms~\cite{maena_gain_cell_aCAM}, and even replacing power-hungry \acp{ADC} in CIM systems~\cite{Zhao2023RACEITAR}.

The memristive device comparator, shown in Figure~\ref{fig:aCAM_intro}~b) and c), is a fundamental building block of most memristive device-based \acp{aCAM} architectures~\cite{Li2020,manea2023non}, as it enables analog inequality checks directly within the memory cell. Two comparator orientations are shown: the \emph{top-connected} configuration in Figure~\ref{fig:aCAM_intro}~b), where the RRAM \ac{OE} is connected to the access transistor, and the \emph{bottom-connected} configuration in Figure~\ref{fig:aCAM_intro}~c), where the \ac{AE} is connected to the access transistor. The comparison threshold is encoded in the memristive device conductance, allowing each cell to implement a programmable analog decision boundary. Multiple boundaries, corresponding to distinct conductance states, can be realized, as illustrated by the input sweep in Figure~\ref{fig:aCAM_intro}~d). We also employ a 2T1R structure where the LD transistor is used for the \ac{aCAM} operation as well as the SET; the HD transistor is exclusively used for RESET.

The chip integrates two $8 \times 8$ \ac{aCAM} arrays (CAM1/CAM2). Each array is not composed of a single uniform cell type; instead, different rows implement different \ac{aCAM} cell architectures, including the compact 6T2M design~\cite{Li2020}, the improved 10T2M variant~\cite{Bazzi2022}, and an enhanced architecture developed specifically for this tape-out~\cite{manea2023non}. This row-wise organization enables several cell variants to be evaluated within the limited available chip area, while keeping them electrically separable during measurements. 

In addition, two distinct \ac{ML} configurations are implemented: a NOR-type and a NAND-type match line. Each row of \ac{aCAM} cells is connected to both \acp{ML} in parallel, allowing the same stored cell states to be evaluated simultaneously by the two match-line schemes. This enables a direct comparison of speed, energy consumption, and functional behavior.

The two $8 \times 8$ \ac{aCAM} arrays implement the two RRAM comparator orientations shown in Figure~\ref{fig:aCAM_intro}~b) and c): CAM1 uses the \emph{top-connected} configuration shown in Figure~\ref{fig:aCAM_intro}~b), where the RRAM \ac{OE} is connected to the access transistor, whereas CAM2 uses the \emph{bottom-connected} configuration shown in Figure~\ref{fig:aCAM_intro}~c), where the \ac{AE} is connected to the access transistor. These configurations exhibit different trade-offs due to polarity-dependent switching in filamentary VCM RRAMs and transistor body effects~\cite{Dittmann2022001,10461986,bengel2023tailor}. With \ac{AE}-to-transistor connection, the body effect appears during RESET and can limit the achievable LCS, whereas \ac{OE}-to-transistor connection shifts it to SET, potentially degrading programming precision and linearity. Since the RRAM is intentionally read in the RESET direction to reduce read disturb and improve retention~\cite{electronics13132639,Dittmann2022001}, the electrode orientation also affects \ac{aCAM} read behavior. In particular, the bottom-connected configuration in CAM2 may introduce body effects during readout, potentially reducing the read margin. The resulting \ac{aCAM}-specific trade-offs between CAM1 and CAM2 are investigated through simulation and evaluated in Section~\ref{sec:res_cam}.

\subsection{CIM arrays}
\begin{figure}[htb!]
    \centering
    \includegraphics[width=1\linewidth]{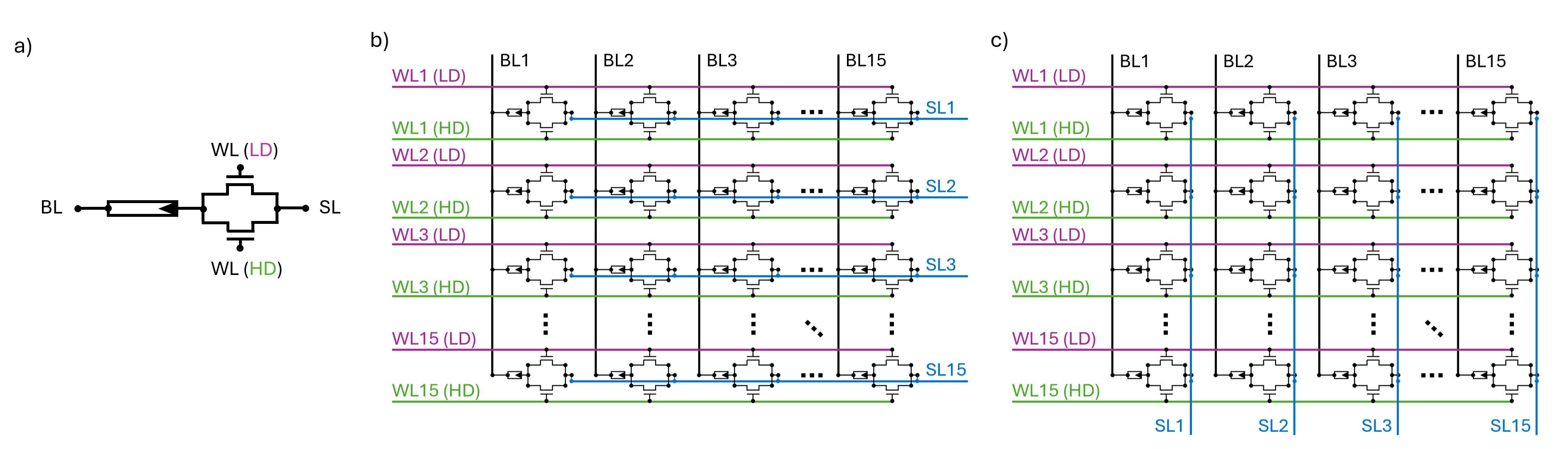}
    \caption{Schematic of the 2T1R cell structures and two CIM array designs implemented on \textit{OTTER}. (a) Single 2T1R cell combining one HD and one LD transistor in parallel. (b) CIM1: Pseudo-crossbar array enabling analog VMM. (c) CIM2: Conventional array for memory application with ideal selectivity. }
    \label{fig:CIMdesign}
\end{figure}

CIM arrays enable arithmetic operations directly within memory arrays, thereby reducing data transfer between memory and processing units compared to conventional von Neumann architectures. This approach is particularly attractive for machine learning workloads, where VMM and multiply-accumulate (MAC) operations dominate the computational complexity. In memristive CIM systems, RRAM conductance states encode matrix weights, allowing analog MAC operations to be performed intrinsically within the array using Ohm’s and Kirchhoff’s laws.\\
\textit{OTTER} integrates two CIM arrays, both based on the proposed 2T1R cell structure, but employing different array organizations to explore trade-offs between reliable programming and computational efficiency. The first array is depicted in Figure~\ref{fig:CIMdesign}~b) and adopts a pseudo-crossbar design, in which the word lines are arranged in parallel with the source lines. While this layout complicates programming due to reduced selectivity and increased interaction between cells, it is better suited for compute-in-memory operations. In particular, the pseudo-crossbar structure enables analog vector-matrix multiplication (VMM) by naturally exploiting current summation along shared lines, a functionality that cannot be efficiently realized in conventional memory arrays.

The second array, sketched in Figure~\ref{fig:CIMdesign}~c) follows a conventional architecture analogous to standard 1T1R memory arrays, where bit lines (BLs) and source lines (SLs) run in parallel, while word lines (WLs) are oriented perpendicularly. This configuration enables precise selection of individual memory cells and provides robust and well-controlled programming conditions, making it suitable for memory-oriented operation and reference measurements.
Both arrays span $15 \times 15$ 2T1R structures and are designed to allow a direct comparison between conventional memory-oriented operation and computation-optimized architectures within the same technology platform.

\section{TaO\textsubscript{x} based VCM device integration on 28 nm CMOS chip}
\begin{figure*}[tbh]
    \centering
    \includegraphics[width=0.85\textwidth]{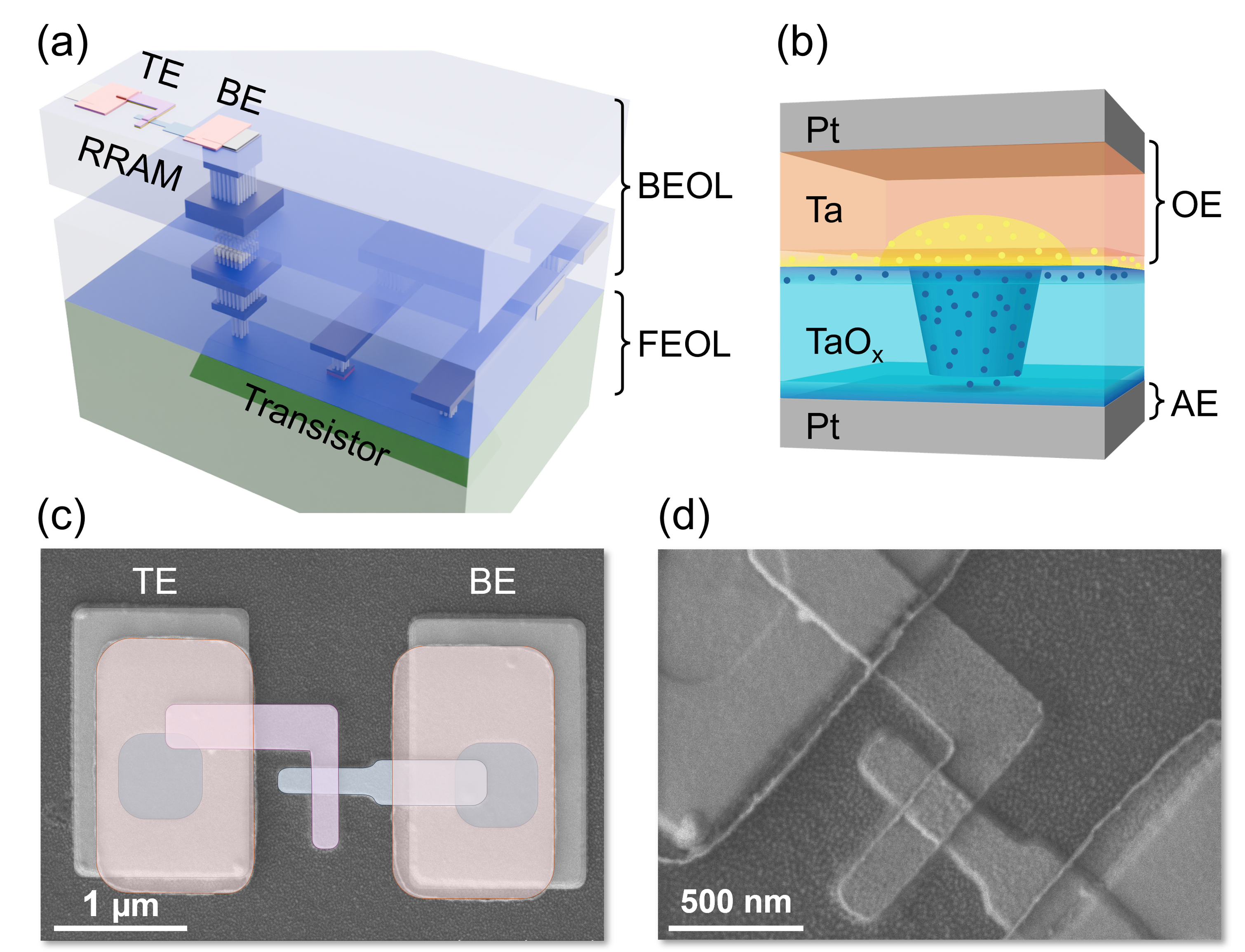}
    \caption{
    (a) Drawing of the nano-crossbar RRAM device integrated on top of the CMOS chip with the active electrode, i.e. the bottom electrode (BE), connected to the source contact of the front end foundry transistor. (b) RRAM stack in the crossbar junction, consisting (from the bottom to the top) of a 25\,nm thick Pt active electrode (AE; grey), a 7\,nm thick TaO\textsubscript{x} (x\,$<$\,2.5) switching layer (blue),  a 13\,nm thick Ta ohmic electrode (OE; orange), and a 25\,nm Pt capping layer (grey). The switching occurs in the area of the conductive filament that is defined by the conductive plug of reduced TaO\textsubscript{y} (y\,$<$\,x ;dark blue) and the disc area. The local reduction of the average Ta valence from +2x to +2y is caused by 
     oxygen exchange with the reactive Ta electrode indicated by the yellow color.
    (c) and (d) Scanning electron microscope images of a 200\,nm\,$\times$\,200\,nm RRAM device integrated on the CMOS chip. The BE and TE electrodes are connected to the transistor and a measurement pad, respectively. The color code in (a) and (c) is opening of the Cu-pads: dark blue, CPs: grey, BE: blue, TE: violet, CBs: light brown.}
    \label{fig:co-integration}
\end{figure*}

The RRAM devices investigated were integrated on top of a custom-designed silicon chip in 28\,nm  technology from TSMC Ltd. with a die size of 36\,mm$^2$ (see Figure~\ref{fig:d0design}), using processes compatible with CMOS back-end-of-line (BEOL) manufacturing. The chip layout contains transistor structures that allow a fabrication of single 1T1R and 2T1R cells as well as arrays of $15 \times 15$ 2T1R structures which enable CIM and CAM operations as discussed in Section~\ref{sec:Design}. The NMOS transistors are rated for up to 3.3\,V and have dimensions of $W/L$\,=\,2 (LD) and $W/L$\,=\,10 (HD). The $200\,\si{\nano\meter} \times 200\,\si{\nano\meter}$--sized Pt/7\,nm TaO\textsubscript{x}/13\,nm Ta/Pt memristive devices in crossbar structure were fabricated in approximately thirty BEOL--compatible processing steps at temperatures below 300\,$\si{\degree}$C using CMOS--compatible materials and chemicals under state-of-the-art clean room conditions in the Helmholtz Nanofacility~\cite{HelmholtzNanoFacility}. A 
detailed description of each fabrication step is provided in the Supplementary Information. The manufacturing process involves four major elements: the contact pads (CP),
the RRAM bottom electrode (BE), the RRAM switching layer and top electrode (TE),
and the connector bars (CB), which are shown in Figure~\ref{fig:co-integration}\,a,c)
in different colors. The process starts with the thinning of the SiO\textsubscript{2}
passivation layer and the selective opening of the Cu pads in the M10 layer, which
are immediately covered with 5\,nm~Ta and 50\,nm~Pt as corrosion protection.

The
$2.0\,\si{\micro\meter} \times 1.3\,\si{\micro\meter}$ CPs for the RRAM
(see Figure~\ref{fig:co-integration}\,c)) and the
$75\,\si{\micro\meter} \times 75\,\si{\micro\meter}$ CPs for the probe pins to the
underlying chip (see Figure~\ref{fig:d0design}) are subsequently defined by electron
beam lithography (EBL) and Ar-based reactive ion beam etching (RIBE) of the Ta/Pt
film. The BE is formed from a sputtered stack of 5\,nm~Ta and 25\,nm~Pt, patterned
by the same procedure, which is intended to ensure a clean interface between the
Pt~AE and the switching layer. In the third step, 7\,nm of TaO\textsubscript{x},
13\,nm of Ta and 25\,nm of Pt are deposited by reactive and inert sputtering and
structured in the shape of the TE. In the final step, the RRAM electrodes are
connected to the respective CPs by means of CBs of
$1.0\,\si{\micro\meter} \times 1.7\,\si{\micro\meter}$ size, fabricated by lift-off.
A detailed description of each fabrication step, including all lithography, etching
and resist-stripping parameters, is given in Section~S1 of the Supplementary
Information; the chemical compatibility studies that determined the choice of
photoresists and resist strippers are documented in Section~S2.

\section{Experimental}
\label{sec:Experimental}

\subsection{2T1R Characterization}
\begin{figure}[tbh]
    \centering
    \includegraphics[width=0.9\linewidth]{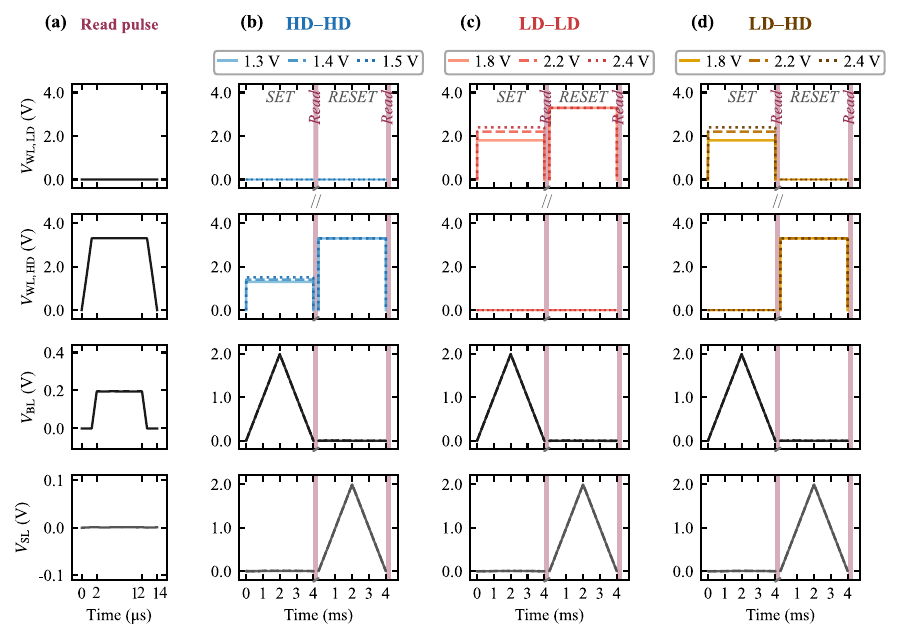}
    \caption{Applied voltage waveforms for the three 2T1R transistor-pairing configurations.
(a)~Read pulse ($V_{\mathrm{BL}} = 0.2\,\mathrm{V}$, $V_{\mathrm{WL,HD}} = 3.3\,\mathrm{V}$,
$t = 10\,\text{\textmu s}$, applied after each SET and RESET operation.
(b)~HD--HD: both SET and RESET are driven by the HD transistor ($W/L = 10$),
with $V_{\mathrm{WL,HD}} \in \{1.3, 1.4, 1.5\}\,\mathrm{V}$ during SET
and $V_{\mathrm{WL,HD}} = 3.3\,\mathrm{V}$ during RESET.
(c)~LD--LD: both SET and RESET are driven by the LD transistor ($W/L = 2$),
with $V_{\mathrm{WL,LD}} \in \{1.8, 2.2, 2.4\}\,\mathrm{V}$ during SET
and $V_{\mathrm{WL,LD}} = 3.3\,\mathrm{V}$ during RESET.
(d)~LD--HD: SET is driven by the LD transistor
($V_{\mathrm{WL,LD}} \in \{1.8, 2.2, 2.4\}\,\mathrm{V}$),
while RESET is driven by the HD transistor ($V_{\mathrm{WL,HD}} = 3.3\,\mathrm{V}$).
In all configurations, a triangular voltage sweep
($0\,\mathrm{V} \to 2\,\mathrm{V} \to 0\,\mathrm{V}$, $4\,\mathrm{ms}$ total)
is applied to $V_{\mathrm{BL}}$ during SET and to $V_{\mathrm{SL}}$ during RESET.
Shaded regions indicate the timing of the read pulses.
For clarity, the RESET phase is displayed with an independent time axis starting from zero,
with the initial label omitted to avoid overlap.
The read pulse between SET and RESET is not shown to scale.}
    \label{fig:Waveform_2T1R}
\end{figure}

The 2T1R architecture incorporates two transistors sharing a common RRAM device and therefore requires five probe contacts: the bit line ($V_{\mathrm{BL}}$), the source line 
($V_{\mathrm{SL}}$), the WL of the LD and HD transistors ($V_{\mathrm{WL,LD}}$ and $V_{\mathrm{WL,HD}}$), and the transistor bulk terminal. In all 2T1R measurements, the bulk 
terminal was connected via a dedicated probe needle to the chassis ground of the Keithley 4200A-SCS, ensuring a well-defined body potential throughout the switching characterization. Note that during RESET the effective source of the transistor floats with the voltage divider formed by the RRAM device and the transistor channel, so that a body effect is present under these conditions; its influence on the RESET behavior is analyzed 
in Section~\ref{sec:design_guidelines}.
Three transistor-pairing configurations were systematically investigated within the 2T1R architecture shown in Figure~\ref{fig:Waveform_2T1R}. In all configurations, a read pulse ($V_\mathrm{BL} = 0.2\,\mathrm{V}$, $V_\mathrm{WL,HD} = 3.3\,\mathrm{V}$, $t = 10$\,\textmu s) was inserted after each SET operation and after each RESET operation to record the instantaneous resistance state (see Figure~\ref{fig:Waveform_2T1R}~a)). 

The HD transistor was used exclusively for readout across all configurations to ensure a consistent and comparable read condition. The read current was extracted as the mean of the central 60\% of the data points acquired during each read pulse, excluding the leading and trailing edges to eliminate the influence of capacitive charging and discharging transients.

In the following, the notation X--Y denotes the transistor geometry used for SET (X) and RESET (Y) operations, respectively. In the HD--HD configuration shown in Figure~\ref{fig:Waveform_2T1R}~b), both SET and RESET were driven by the HD transistor, with $V_\mathrm{WL,HD} \in \{1.3, 1.4, 1.5\}\,\mathrm{V}$ during SET and $V_\mathrm{WL,HD} = 3.3\,\mathrm{V}$ during RESET. In the LD--LD configuration (Figure~\ref{fig:Waveform_2T1R}~c)), both operations were driven by the LD transistor, with $V_\mathrm{WL,LD} \in \{1.8, 2.2, 2.4\}\,\mathrm{V}$ during SET and $V_\mathrm{WL,LD} = 3.3\,\mathrm{V}$ during RESET. In the LD--HD configuration shown in Figure~\ref{fig:Waveform_2T1R}~d), SET was performed by the LD transistor ($V_\mathrm{WL,LD} \in \{1.8, 2.2, 2.4\}\,\mathrm{V}$) and RESET by the HD transistor ($V_\mathrm{WL,HD} = 3.3\,\mathrm{V}$). 
In all cases, a triangular voltage sweep ($0\,\mathrm{V} \to 2\,\mathrm{V} \to 0\,\mathrm{V}$, $4\,\mathrm{ms}$ total) was applied to $V_\mathrm{BL}$ during SET and to $V_\mathrm{SL}$ during RESET. The applied waveforms are illustrated in Figure~\ref{fig:Waveform_2T1R}, and the resulting current--voltage characteristics are discussed in Section~\ref{sec:Results} with reference to Figure~\ref{fig:2t1r_HDHD_LDLD_LDHD}.

\begin{figure}[tbh]
    \centering
    \includegraphics[width=0.9\linewidth]{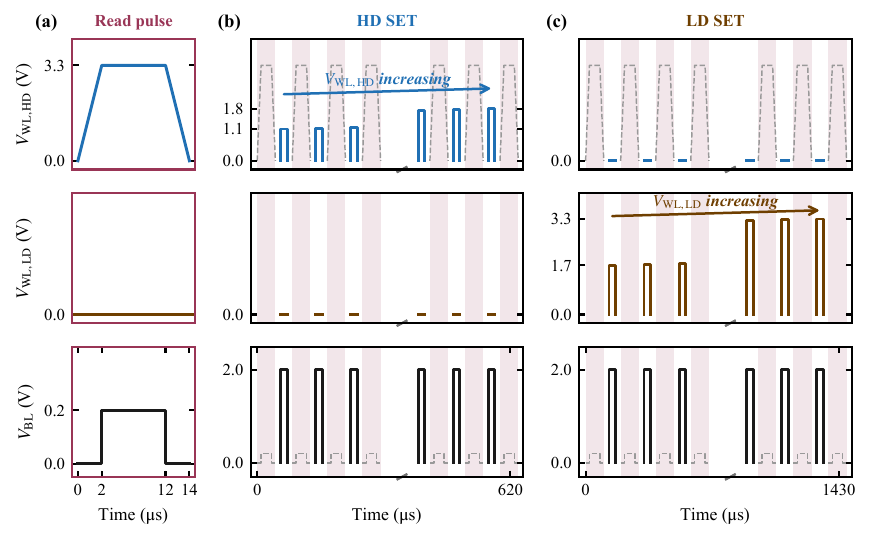}
    \caption{Schematic illustration of the pulse training protocol applied to the 2T1R cell following the incremental gate voltage (IGVVA) algorithm. (a)~Read pulse waveform: the HD transistor gate is biased at $V_\mathrm{WL,HD} = 3.3\,\mathrm{V}$ with $V_\mathrm{BL} = 0.2\,\mathrm{V}$ for $10$\,\textmu s to record the instantaneous resistance state. (b)~HD SET protocol: $V_\mathrm{WL,HD}$ is incremented from $1.1\,\mathrm{V}$ to $1.8\,\mathrm{V}$ in steps of $0.025\,\mathrm{V}$ across 29 SET pulses ($1$\,\textmu s each), with $V_\mathrm{WL,LD} = 0$\,V throughout. (c)~LD SET protocol: $V_\mathrm{WL,LD}$ is incremented from $1.7\,\mathrm{V}$ to $3.3\,\mathrm{V}$ in steps of $0.025\,\mathrm{V}$ across 65 SET pulses, with $V_\mathrm{WL,HD} = 0$\,V throughout. In both protocols, a read pulse (shaded regions; see panel~(a)) is inserted after each SET pulse to record the resistance state, and a final RESET sweep driven by the HD transistor restores the high-resistance initial state prior to the next programming cycle. Only the first and last three pulses are shown; intermediate pulses are omitted for clarity. The time axis is therefore not drawn to scale.}
    \label{fig:Pulse_training_waveform}
\end{figure}

The three transistor-pairing configurations characterized above establish the feasibility of the LD--HD architecture and reveal the complementary roles of the two transistor geometries in resistive switching. However, pulse-based programming more faithfully represents the operating conditions of neuromorphic computing circuits, where synaptic weights are updated through sequences of fixed-duration pulses. To evaluate the analog programmability of the 2T1R cell under such conditions, a pulse training characterization was conducted using either the HD and LD transistors for SET, following the incremental gate voltage verify algorithm (IGVVA)~\cite{miloOptimizedProgrammingAlgorithms2021,perezVariabilityEnergyConsumption2021}, with RESET performed exclusively by the HD transistor in both cases shown in Figure~\ref{fig:Pulse_training_waveform}. The resulting programmability and variability characteristics are presented in Section~\ref{sec:Results}, with reference to Figures~\ref{fig:2t1r_pulse}, ~\ref{fig:2t1r_MLC}, and~\ref{fig:2t1r_algo}. 
Figures~\ref{fig:Pulse_training_waveform}\,~b) and~c) illustrate the HD and LD SET protocols, corresponding to an HD--HD and LD--HD operation, respectively. In the HD SET protocol, $V_{\mathrm{WL,HD}}$ is incremented from $1.1\,\mathrm{V}$ to $1.8\,\mathrm{V}$ in steps of $0.025\,\mathrm{V}$ across 29 pulses; in the LD SET protocol, $V_\mathrm{WL,LD}$ is incremented from $1.7\,\mathrm{V}$ to $3.3\,\mathrm{V}$ across 65 pulses, with the upper limit defined by the nominal operating voltage of the 28\,nm process node. In both protocols, each SET pulse is $1$\,\textmu s in duration and is followed by a read pulse ($V_\mathrm{WL,HD} = 3.3\,\mathrm{V}$, $V_\mathrm{BL} = 0.2\,\mathrm{V}$, $t = 10$\,\textmu s) to record the instantaneous resistance state (shaded regions; see Figure~\ref{fig:Pulse_training_waveform}~a)).

 The read current is extracted as the mean of the central 60\,\% of the data points acquired during each read pulse, excluding the leading and trailing edges to eliminate capacitive charging and discharging transients. The sequence concludes with a RESET sweep driven by the HD transistor to restore the high-resistance initial state.

\section{Compact Model}
\label{sec:Compact Model}
To simulate the switching dynamics of the co-integrated TaO$_x$-based VCM devices, we employ the JART VCM Rth compact model~\cite{sonComprehensiveCompactModel2025}. The key extension over the predecessor JART VCM v1b model~\cite{Bengel969153} is a state-dependent effective thermal resistance:
\begin{equation}
    R_{\text{th,eff}} = \frac{\beta}{\alpha + \gamma N_{\text{disc}}^{2}},
    \label{eq:Rtheff}
\end{equation}
where $N_{\text{disc}}$ is the oxygen vacancy concentration in the disc region and $\alpha$, $\beta$, $\gamma$ are polarity-dependent fitting coefficients. The evolution of $N_{\text{disc}}$ is governed by thermally activated ionic transport:
\begin{equation}
    \frac{\mathrm{d}N_{\text{disc}}}{\mathrm{d}t}
    = -\frac{I_{\text{ion}}}{z_{V_\text{O}} e A l_{\text{disc}}},
    \label{eq:dNdt}
\end{equation}
where $z_{V_\text{O}}$ is the oxygen vacancy charge number, $e$ the elementary charge, $A = \pi R_{\text{fil}}^2$ the filament cross-sectional area with $R_{\text{fil}}$ the filament radius, $l_{\text{disc}}$ the disc length, and $I_{\text{ion}}$ the ionic current describing the thermally activated motion of oxygen vacancies within the filament. The local filament temperature entering this expression is determined by Joule heating via
\begin{equation}
    T = \left(V_{\text{disc}} + V_{\text{plug}} + V_{\text{Schottky}}\right) I \cdot R_{\text{th,eff}} + T_0,
    \label{eq:Tlocal}
\end{equation}
where $V_{\text{disc}}$, $V_{\text{plug}}$, and $V_{\text{Schottky}}$ are the voltage drops across the disc region, the plug region, and the Schottky barrier at the active electrode, respectively, $I$ is the total device current, and $T_0$ is the ambient temperature. This mutual coupling between $N_{\text{disc}}$, $R_{\text{th,eff}}$, and $T$ enables a self-consistent description of gradual multilevel switching. During SET, $R_{\text{th,eff}}$ decreases with increasing $N_{\text{disc}}$, moderating the thermal runaway and supporting stable multilevel programming. During RESET, the opposite trend leads to an increasing $R_{\text{th,eff}}$, resulting in a gradual current reduction consistent with experimental observations. The corresponding reference parameter set is summarised in Table~\ref{tab:jart_params}.

\begin{table}[tb]
\centering
\caption{Reference parameter set for the JART VCM Rth compact model.}
\label{tab:jart_params}
\begin{tabular}{@{}lll@{}}
\hline
\textbf{Symbol} & \textbf{Description} & \textbf{Value} \\
\hline
$l_{\text{cell}}$         & Cell thickness                    & 7\,nm \\
$l_{\text{disc}}$         & Disc region length                & 3\,nm \\
$l_{\text{plug}}$         & Plug region length                & $l_{\text{cell}}-l_{\text{disc}}$ \\
$R_{\text{fil}}$         & Filament radius                   & 30\,nm \\
$z_{V_\mathrm{O}}$        & Vacancy charge number             & 2 \\
$a$                       & Ion hopping distance              & 0.3\,nm \\
$\nu_0$                   & Attempt frequency                 & $2\times10^{12}$\,Hz \\
$\Delta W_{\text{A}}$     & Ionic hopping activation energy   & 0.9\,eV \\
$m_{\text{eff}}$          & Effective electron mass           & $9.1\times10^{-31}$\,kg \\
$e\phi_{\text{Bn0}}$      & Zero-bias Schottky barrier height & 0.5\,eV \\
$e\phi_{\text{n}}$        & Conduction band offset            & 0.1\,eV \\
$\mu_{\text{n}}$          & Electron mobility                 & $7.5\times10^{-6}$\,m$^2$/(Vs) \\
$N_{\text{max}}$          & Maximum vacancy concentration     & $30\times10^{26}$\,m$^{-3}$ \\
$N_{\text{min}}$          & Minimum vacancy concentration     & $0.1\times10^{26}$\,m$^{-3}$ \\
$N_{\text{plug}}$         & Plug vacancy concentration        & $N_{\text{max}}$ \\
$\varepsilon_r$           & Relative permittivity             & 27 \\
$\varepsilon_{r,\phi_B}$  & Barrier region permittivity       & 4 \\
$T_0$                     & Ambient temperature               & 273.15\,K \\
\hline
\end{tabular}
\end{table}

\begin{table}[tb]
\centering
\caption{$R_{\text{th}}$ coefficients and their ranges across all switching configurations and gate bias conditions.}
\label{tab:Rth_params}
\begin{tabular}{@{}llllll@{}}
\hline
\textbf{Symbol} & \textbf{Median} & \textbf{Range} & \textbf{Symbol} & \textbf{Median} & \textbf{Range} \\
\hline
$\alpha_{\text{SET}}$                       & $0.10$  & $[0.055,\,0.27]$  & $\alpha_{\text{RESET}}$                       & $0.45$  & $[0.05,\,0.50]$ \\
$\beta_{\text{SET}}$ ($10^{6}$\,K/W)        & $2.3$   & $[1.0,\,3.0]$     & $\beta_{\text{RESET}}$ ($10^{5}$\,K/W)        & $7.0$   & $[1.15,\,12.0]$ \\
$\gamma_{\text{SET}}$ ($10^{-52}$\,m$^{6}$) & $3.0$   & $[0.45,\,4.8]$    & $\gamma_{\text{RESET}}$ ($10^{-53}$\,m$^{6}$) & $0.5$   & $[0.01,\,1.0]$ \\
\hline
\end{tabular}
\end{table}

The access transistor is incorporated into the circuit-level simulation via a calibrated process design kit (PDK) compact model, with key parameters adjusted to match the measured transfer characteristics of the HD and LD   devices. To model device-to-device variability, $l_{\text{disc}}$, $\Delta W_{\text{A}}$ and $N_{\text{min}}$ are treated as the adjustable parameters reflecting variations in the initial state across devices, while all remaining physical parameters are fixed at the reference values in Table~\ref{tab:jart_params}. Since cycle-to-cycle variability and the varying gate bias conditions across the three switching configurations (HD--HD, LD--LD, and LD--HD) prevent a single fixed parameter set from consistently reproducing the switching behavior, the $R_{\text{th}}$ coefficients are individually calibrated for each configuration and gate voltage condition. The resulting reference values and the total parameter ranges encountered across all conditions are reported in Table~\ref{tab:Rth_params}.

To extend the simulation beyond deterministic sweep fitting and to reproduce the statistical spread of resistance states observed across repeated switching cycles, a variability-aware simulation framework following Ahmad \textit{et al.}~\cite{Ahmad_2024} is employed. In this framework, the parameters $N_{\text{min}}$, $R_{\text{fil}}$, and $l_{\text{new}}$ are updated between consecutive switching cycles via a multiplicative random walk. The random walk is constrained within bounds defined individually for each switching configuration and gate bias condition. The calibration of both the deterministic switching characteristics and the resulting resistance state distributions against experimental data is presented in Section~\ref{sec:Results}.

\section{Results}
\label{sec:Results}
While showcasing \textit{OTTER} as a research vehicle, this paper investigates the 2T1R architecture through systematic experimental characterization accompanied by physical compact simulations. The following sections present DC sweep and pulse-based programming results for the 2T1R cell, from which design guidelines for transistor sizing are derived. Simulation-based CAM design considerations are presented in Section~\ref{sec:res_cam}. Finally, a multiply-and-accumulate operation using CIM\,1 (see Figure~\ref{fig:d0design}) is demonstrated.
\subsection{2T1R Switching Characterization}
This section presents the experimental characterization of the 2T1R structures, progressing from static triangular current--voltage sweep measurements to pulse-based multilevel programming. The results demonstrate the complementary roles of the HD and LD access transistors and establish the advantage of the 2T1R architecture for reliable analog weight programming. The DC sweep characterization is compared with simulated characteristics obtained using the JART VCM Rth compact model, allowing the experimentally observed gradual SET evolution and multilevel switching behavior to be reproduced.

\begin{figure}[tb]
    \centering
    \includegraphics[width=0.9\linewidth]{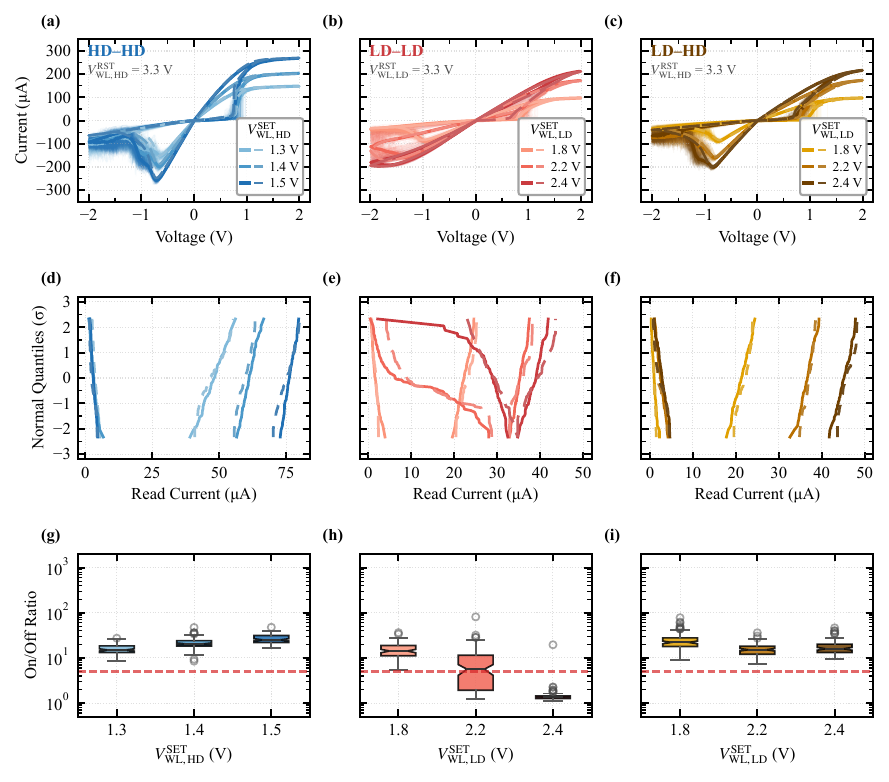}
    \caption{Switching characteristics of the three 2T1R transistor-pairing configurations. (a)--(c)~Current--voltage characteristics for HD--HD, LD--LD, and LD--HD, respectively, measured over 100 cycles. Solid lines represent the cycle-averaged experimental results and dashed lines indicate the corresponding simulation results. (d)--(f)~Distributions of the read current in the LRS and HRS, recorded immediately after SET and RESET, respectively. Annotated values indicate the median read current in \textmu A. (g)--(i)~On/off ratio statistics as a function of SET gate voltage. The dashed red line marks an on/off ratio of 5. In all configurations, the SET operation applies a triangular sweep to $V_\mathrm{BL}$ and RESET to $V_\mathrm{SL}$ ($0\,\mathrm{V} \to 2\,\mathrm{V} \to 0\,\mathrm{V}$, $4\,\mathrm{ms}$ total).}
    \label{fig:2t1r_HDHD_LDLD_LDHD}
\end{figure}

\textit{OTTER} incorporates such a 2T1R structure, characterized following the sweep protocol described in Section~\ref{sec:Experimental} and illustrated in Figure~\ref{fig:Waveform_2T1R}. The resulting characteristics for the three transistor-pairing configurations (HD–HD, LD–LD, and LD–HD) are presented in Figure~\ref{fig:2t1r_HDHD_LDLD_LDHD}.
The current--voltage characteristics are given in panels (a)--(c), the read current distributions in panels (d)--(f), and the on/off ratio statistics in panels (g)--(i). 

The HD--HD configuration, shown in Figures~\ref{fig:2t1r_HDHD_LDLD_LDHD}~a),~d), and~g), in which both SET and RESET are driven by the HD transistor, serves as the reference case. The blue curves from light to dark correspond to SET WL voltages of 1.3\,V, 1.4\,V, and 1.5\,V, respectively, with the HD transistor fully opened during RESET ($V_{\mathrm{WL,HD}} = 3.3\, \mathrm{V}$). Solid lines represent the median current over 100 measurement cycles, and dashed lines indicate the corresponding simulation results. The HD transistor provides effective current compliance during SET, with $I_{\mathrm{cc}}$ increasing monotonically with WL bias and exhibiting good cycle-to-cycle consistency. The distributions in Figure~\ref{fig:2t1r_HDHD_LDLD_LDHD}\,~d) confirm well-separated LRS and HRS populations, and the on/off ratio statistics in Figure~\ref{fig:2t1r_HDHD_LDLD_LDHD}\,~g) show that all cycles consistently exceed the threshold of 5 across all gate bias conditions, demonstrating reliable binary switching behavior.

The LD--LD configuration (Figures~\ref{fig:2t1r_HDHD_LDLD_LDHD}~b), e), and h)), in which both operations are driven by the LD transistor, reveals a strong WL-bias dependence of the RESET performance. At 1.8\,V ($I_\mathrm{cc} \approx 100$\,\textmu A), complete RESET is achieved and on/off ratios consistently exceed 5. As the SET WL voltage increases to 2.2\,V ($I_\mathrm{cc} \approx 175$\,\textmu A), approximately 50\,\% of cycles fail to return to the HRS, with the distribution showing a collapse of the HRS population. At 2.4\,V ($I_\mathrm{cc} \approx 200$\,\textmu A), RESET fails in the vast majority of cycles, and the current--voltage characteristics during RESET follow the transistor output characteristic rather than the RRAM switching characteristic, indicating that the device remains in the LRS due to the series resistance of the LD transistor.

To address this limitation, the LD--HD configuration decouples the SET and RESET operations across the two transistors: SET is performed by the LD transistor and RESET by the HD transistor. The SET WL voltages and device under test are identical to those used in the LD--LD characterization, enabling a direct comparison. As shown in Figures~\ref{fig:2t1r_HDHD_LDLD_LDHD}~c), f), and i), replacing the LD transistor with the HD transistor for the RESET operation fully restores switching reliability across all SET WL bias conditions. Complete RESET is achieved even after using the highest SET WL bias of 2.4\,V, the LRS and HRS distributions remain well separated, and on/off ratios consistently exceed 5 in all cycles.

\begin{figure}[tb]
    \centering
    \includegraphics[width=0.9\linewidth]{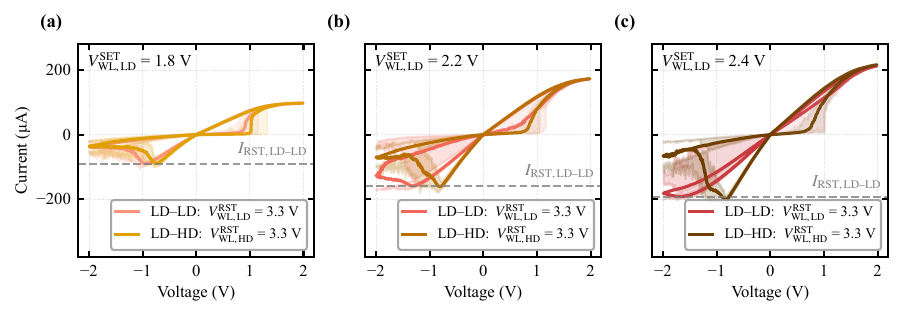}
    \caption{Current--voltage characteristics of the LD--LD and LD--HD configurations at SET WL voltages of (a)~$1.8\,\mathrm{V}$, (b)~$2.2\,\mathrm{V}$, and (c)~$2.4\,\mathrm{V}$. In both configurations, SET is driven by the LD transistor with an identical BL sweep; RESET is driven by the LD transistor ($V_\mathrm{WL,LD}^\mathrm{RST} = 3.3\,\mathrm{V}$) in LD--LD and by the HD transistor ($V_\mathrm{WL,HD}^\mathrm{RST} = 3.3\,\mathrm{V}$) in LD--HD. Shaded regions indicate cycle-to-cycle variability. The dashed line marks the minimum RESET current observed in the LD--LD configuration ($I_{\mathrm{RST,LDLD}}$) and is included as a reference.}
    \label{fig:2t1r_compare_LDLD_LDHD}
\end{figure}

Figure~\ref{fig:2t1r_compare_LDLD_LDHD} directly compares the current--voltage characteristics of the LD--LD and LD--HD configurations under identical SET conditions, isolating the effect of the RESET transistor geometry.
In both configurations, SET is driven by the LD transistor at the same WL voltages; the only difference is that RESET is performed by the LD transistor in the LD--LD case and by the HD transistor in the LD--HD case. The dashed grey line in each panel marks the minimum RESET current of the LD--LD configuration ($I_\mathrm{RST,LDLD}$), confirming that the peak RESET current reached during the RESET operation is comparable between the two configurations at each SET WL-bias condition. This indicates that any difference in RESET completeness between the two configurations can be attributed solely to the voltage drop across the access transistor, rather than to a difference in drive current.

At $V_{\mathrm{WL,LD}}^{\mathrm{SET}} = 1.8\,\mathrm{V}$, both configurations achieve complete RESET; however, the LD--LD configuration requires a higher SL voltage to initiate RESET compared to the LD--HD configuration. This indicates that the series resistance of the LD transistor is visible across all gate bias conditions, and becomes the dominant limiting factor as the SET current increases. As $V_\mathrm{WL,LD}^\mathrm{SET}$ is increased to $2.2\,\mathrm{V}$ (Figure~\ref{fig:2t1r_compare_LDLD_LDHD}~b)), the RESET becomes incomplete in the LD--LD configuration. Although the RESET current magnitude is comparable to that of the LD--HD case, the excessive voltage drop across the LD transistor prevents a sufficient fraction of $V_\mathrm{SL}$ from being applied across the RRAM element, resulting in only partial RESET. The LD--HD configuration, by contrast, achieves reliable and complete RESET under the same SET conditions. At $V_\mathrm{WL,LD}^\mathrm{SET} = 2.4\,\mathrm{V}$, RESET fails entirely in the LD--LD configuration: the peak RESET current reached by the LD transistor is still comparable to that of the HD transistor, yet the RRAM element cannot be restored to the high-resistance state. The limitation is therefore not the peak current the transistor is able to deliver, but the fraction of $V_\mathrm{SL}$ that remains available across the RRAM element at that current level. The larger on-resistance of the LD transistor shifts the operating point towards a lower voltage across the RRAM device, so that the conditions required for complete RESET are not reached; the underlying transistor-level mechanism is analyzed in
Section~\ref{sec:design_guidelines}. The LD--HD configuration, by contrast, maintains a
complete and reliable RESET under the same SET conditions, demonstrating that the 2T1R architecture effectively decouples the SET current compliance from the RESET drive requirement.

\subsection{Design Guidelines for Transistor Sizing}
\label{sec:design_guidelines}
\begin{figure}[tbh]
    \centering
    \includegraphics[width=0.9\linewidth]{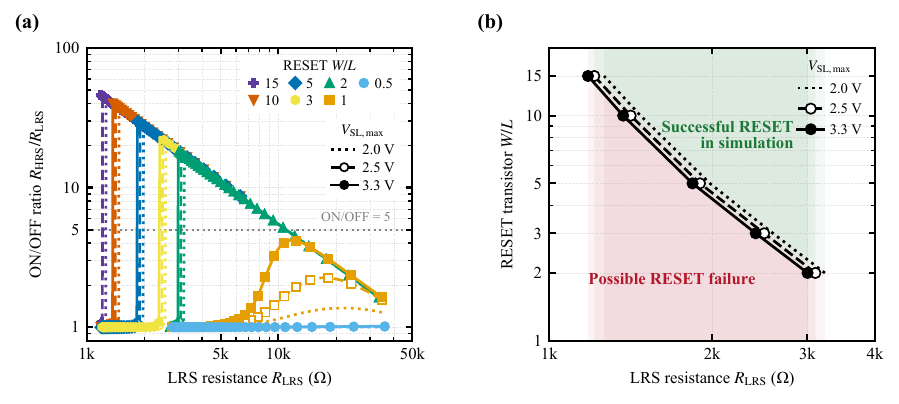}
    \caption{Design guidelines for RESET transistor sizing in the 2T1R cell under different SET current-compliance conditions. (a) Simulated on/off ratio versus programmed LRS for different RESET transistor $W/L$ ratios and $V_{\mathrm{SL,max}}$. The dotted line marks the on/off ratio threshold of 5. (b) Minimum RESET transistor $W/L$ required for an on/off ratio above 5 at different $V_{\mathrm{SL,max}}$.}
    \label{fig:fig_combined_abc}
\end{figure}

The DC sweep characterization in Section~6.1 demonstrates that the JART VCM Rth compact model reproduces the experimentally observed switching behavior across all three transistor-pairing configurations, as shown by the agreement between the measured and simulated curves in Figures~\ref{fig:2t1r_HDHD_LDLD_LDHD}~a)--f). Building on this validation, the compact model is used to extend the analysis beyond the two transistor geometries available on chip, exploring the impact of RESET transistor sizing on switching reliability.\\
In the following simulation, the LRS is programmed under different SET current-compliance levels ($I_{\mathrm{cc}}$), and the RESET transistor $W/L$ ratio is varied across 0.5, 1, 2, 3, 5, 10, and 15. In addition, three maximum SL voltages are considered ($V_\mathrm{SL,max} = 2.0$\,V, 2.5\,V, and 3.3\,V) to evaluate the interplay between RESET voltage headroom and transistor sizing. For each combination of RESET transistor geometry and $V_\mathrm{SL,max}$, the 2T1R cell is programmed to a broad range of LRS values and subsequently RESET.
Figure~\ref{fig:fig_combined_abc}~a) presents the resulting on/off ratio as a function of the programmed LRS for each RESET transistor $W/L$ ratio. The dotted line marks the on/off ratio threshold of 5, consistent with the criterion used in Section~6.1. For small RESET transistor $W/L$ ratios, the on/off ratio drops below the threshold at moderate LRS values, indicating incomplete RESET. As the RESET transistor $W/L$ ratio increases, the on/off ratio remains above the threshold over a progressively wider LRS range. From Figure~\ref{fig:fig_combined_abc}~a), the minimum evaluated RESET transistor $W/L$ ratio satisfying an on/off ratio above 5 can therefore be extracted for each programmed LRS. Figure~\ref{fig:fig_combined_abc}~b) summarizes this minimum $W/L$ ratio for the three $V_{\mathrm{SL,max}}$ conditions, defining the boundary between reliable and incomplete RESET operation. Within the simulated range, this boundary appears to follow a negative double logarithmic trend, which means that a decrease in LRS resistance by one order of magnitude would require an increase in $W/L$ ratio by one order of magnitude to ensure reliable RESET.

The RESET requirement strongly depends on the programmed LRS. For relatively shallow LRS states, corresponding to higher $R_{\mathrm{LRS}}$, RESET can be achieved with a relatively small transistor $W/L$, and the influence of $V_{\mathrm{SL,max}}$ is limited because sufficient RESET capability is already available. As the LRS becomes deeper, increasing $V_{\mathrm{SL,max}}$ can extend the range over which complete RESET is maintained and reduce the required RESET transistor size. For very deep LRS states, however, the benefit of increasing $V_{\mathrm{SL,max}}$ becomes limited, and a sufficiently large RESET transistor $W/L$ is required.

We attribute the diminishing benefit of $V_\mathrm{SL,max}$ at deep LRS states to the operating point of the RESET transistor itself. As $R_\mathrm{LRS}$ decreases, the voltage $V_\mathrm{DS}$ dropping across the channel of the RESET transistor
increases, which drives the transistor from the linear region toward saturation.
Once pinch-off is approached, additional applied RESET voltage is expected to
produce only a limited increase in the voltage and current available to the RRAM
device. For deeply programmed LRS states, increasing $V_\mathrm{SL,max}$ alone
would then be insufficient to trigger RESET, and a larger RESET transistor $W/L$
ratio would be required to provide sufficient RESET current. This interpretation
is consistent with the pinch-off-limited RESET behavior reported by
Zheng~\textit{et~al.}~\cite{Zhengovercoming,}.
\\
Additionally, the body effect is expected to contribute. During RESET, the SL node
is biased whereas the BL node is kept at ground (c.f.\ Figure~\ref{fig:CIMdesign}~a)).
In this case, the effective source of the transistor is at the drain contact
between the RRAM cell and the channel. The potential of this node varies with the
voltage divider formed by the RRAM device and the transistor. The more voltage
drops across the RRAM cell, the higher the potential of the effective source and
its deviation from the bulk potential. As a result, the effective gate--source
voltage $V_\mathrm{GS,effective}$ decreases. This effect is small as long as the
RRAM device is low-resistive, but as soon as the RESET process is triggered and the
RRAM resistance increases, $V_\mathrm{GS,effective}$ starts decreasing, which can
move the system into saturation at rather low current levels. Depending on sweep
rate, switching kinetics and transistor geometry, this mechanism could in principle
lead to a freezing RESET transition. A larger $W/L$ ratio provides more headroom in
current supply, so that a decrease in $V_\mathrm{GS,effective}$ does not shift the
system into saturation.
\\
 It should be noted that these simulations provide indicative trends and qualitative explanations of the links between LRS state ($R_{\mathrm{LRS}}$), transistor $W/L$ ratio and RESET kinetics. The absolute numbers should be interpreted with caution. A direct experimental verification of the proposed
transistor-level mechanisms requires resolving the $I_\mathrm{DS}$--$V_\mathrm{DS}$
trajectory of the RESET transistor during the switching event itself, which is the
subject of ongoing work. Nevertheless, these results establish a general design guideline for transistor sizing in RRAM access structures. For shallow LRS states, RESET can be achieved with a relatively small transistor $W/L$ ratio, while increasing $V_{\mathrm{SL,max}}$ can improve RESET capability at intermediate LRS levels. For deep LRS states, however, the benefit of additional RESET voltage becomes limited due to transistor saturation, and a larger RESET transistor $W/L$ is therefore required to maintain sufficient drive current. Additionally, the body effect becomes more severe for smaller $W/L$ ratios. These considerations are relevant to both 1T1R and 2T1R configurations. In a 1T1R cell, a single transistor must simultaneously provide accurate current compliance during SET and sufficient drive capability during RESET, which restricts the available programming range. In contrast, the 2T1R architecture separates these functions, allowing the SET current-compliance requirement and the RESET transistor sizing to be optimized independently.

\subsection{Pulse-Based Multilevel Programming}
\begin{figure}[tbh]
    \centering
    \includegraphics[width=0.9\linewidth]{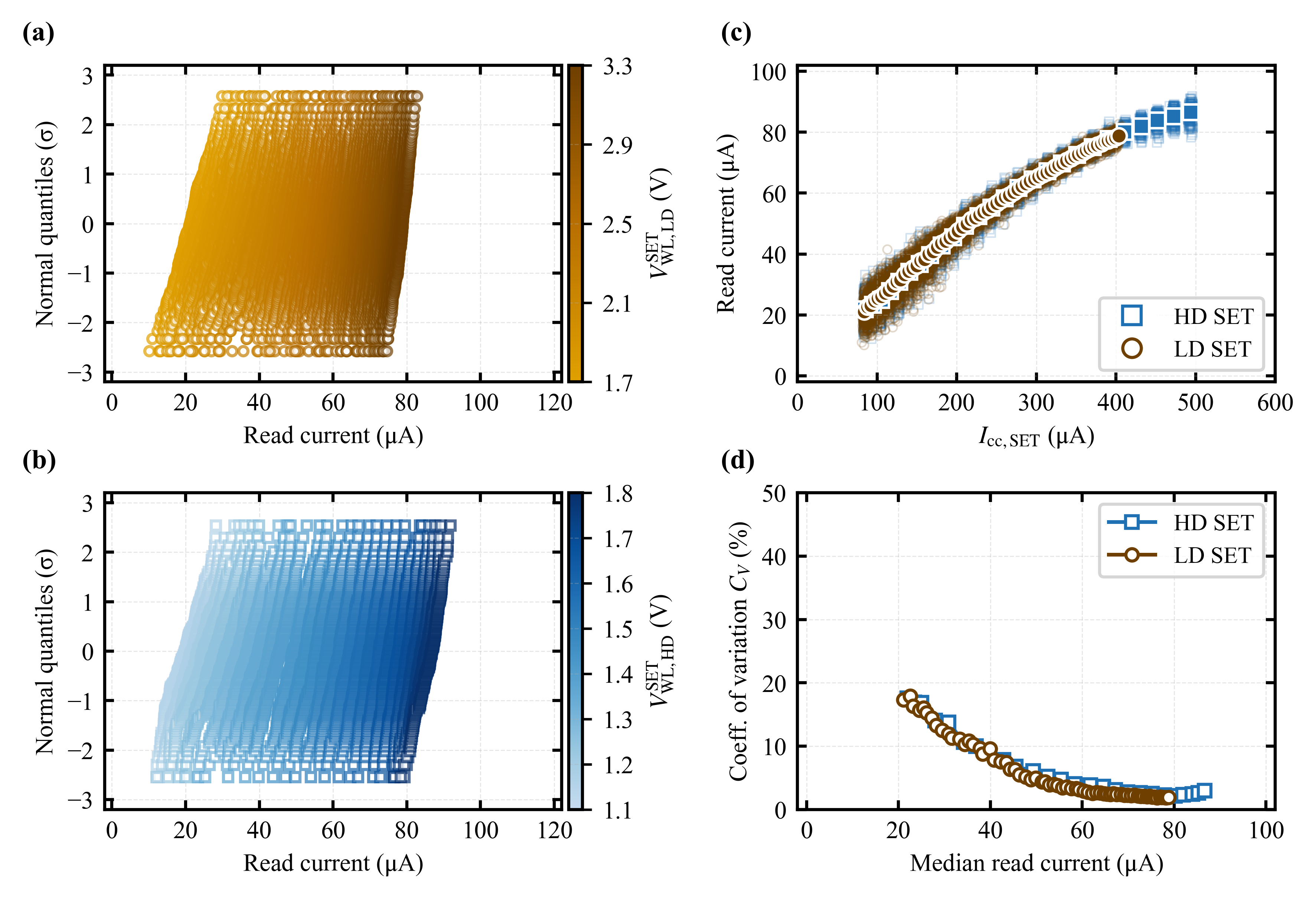}
    \caption{Statistical analysis of the read current distributions programmed through pulse-based multilevel programming of the 2T1R cell under HD SET and LD SET protocols. (a)~Distributions of the read current for the LD SET protocol, with the color gradient from light to dark corresponding to incrementally increasing $V_\mathrm{WL,LD}^\mathrm{SET}$ from $1.7\,\mathrm{V}$ to $3.3\,\mathrm{V}$ in steps of $0.025\,\mathrm{V}$ (65 pulses). (b)~Corresponding distributions for the HD SET protocol, with $V_\mathrm{WL,HD}^\mathrm{SET}$ incremented from $1.1\,\mathrm{V}$ to $1.8\,\mathrm{V}$ (29 pulses). In both protocols, RESET is performed exclusively by the HD transistor. (c)~Read current as a function of SET compliance current $I_\mathrm{cc}$, extracted as the mean of the central 60\% of SL current samples acquired during each SET pulse. (d)~Coefficient of variation $C_V$ of the read current as a function of median read current, quantifying the cycle-to-cycle variability at each programming level. In all panels, blue squares and brown circles denote the HD SET and LD SET protocols, respectively.}
    \label{fig:2t1r_pulse}
\end{figure}

The preceding characterization establishes that the HD transistor provides reliable RESET across all WL-bias conditions, while both transistors support consistent SET operation. However, a key question remains: whether the availability of the LD transistor during SET can improve programming precision under pulse-based operating conditions compared with the HD transistor alone. The 2T1R architecture offers a unique opportunity to address this question directly, as the SET transistor can be selected independently of the RESET transistor on the same device. To this end, the characterization is extended from quasi-static sweep measurements to pulse-based multilevel programming, in which SET is performed using either the HD or the LD transistor while RESET is driven exclusively by the HD transistor in both cases, following the IGVVA. The applied voltage waveforms for both protocols are illustrated in Figure~\ref{fig:Pulse_training_waveform}, and the resulting resistance states after consecutive SET pulses with incrementally increasing $V_{\mathrm{WL}}^{\mathrm{SET}}$ are shown in Figure~\ref{fig:2t1r_pulse}.
Figures~\ref{fig:2t1r_pulse}~a) and~b) show the distributions of the read current for the LD SET and HD SET protocols, respectively, where the color gradient from light to dark corresponds to incrementally increasing SET WL voltage. A comparison of the two panels reveals that for equivalent WL voltage increments of $25\,\mathrm{mV}$, the LD SET protocol produces a significantly denser spacing of the read current distributions than the HD SET protocol. This is a direct consequence of the weaker drive capability of the LD transistor: a given increment in WL voltage produces a smaller change in $I_\mathrm{cc}$, translating into a finer step in the programmed resistance state. As a result, the LD SET protocol yields 65 distinguishable programming levels across a read current range of approximately 10\,\textmu A to 85\,\textmu A, compared to 29 levels for the HD SET protocol over a comparable 
range. Figure~\ref{fig:2t1r_pulse}~c) shows the relationship between the read current and the SET compliance current $I_\mathrm{cc}$, defined as the mean of the central 60\,\% of the source line current samples acquired during each SET pulse. The results for the HD and LD SET protocols are represented by blue squares and brown circles, respectively. Both datasets exhibit a strong and monotonic relationship between $I_\mathrm{cc}$ and the resulting read current, and the two curves overlap closely across the full current range. This confirms that the programmed resistance state is largely determined by $I_\mathrm{cc}$, independent of which transistor drives the SET operation. Consequently, the finer programming granularity of the LD SET protocol is not a result of superior current control precision, but rather of the smaller $I_\mathrm{cc}$ increment per gate voltage step afforded by the narrower transistor geometry. Figure~\ref{fig:2t1r_pulse}\,~d) presents the coefficient of variation $C_V$ of the read current as a function of the median read current at each programming step, providing a direct measure of cycle-to-cycle variability. The HD and LD SET results again overlap closely, confirming that both transistor geometries produce comparable variability characteristics across the full programming range. This demonstrates that the improved programming granularity of the LD SET protocol is achieved without any penalty in variability. Furthermore, $C_V$ decreases monotonically with increasing median read current in both cases, indicating that states deeper in the LRS exhibit lower relative variability, consistent with previously reported behavior in filamentary RRAM devices~\cite{liuEffectPulseSchemes2025,wiefelsReliabilityAspectsResistively}. 

\begin{figure}[tb]
    \centering
    \includegraphics[width=0.9\linewidth]{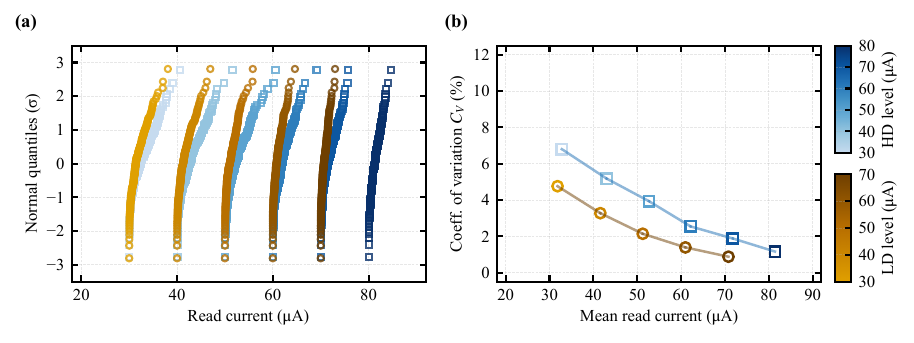}
    \caption{Read current distributions and variability of six target levels extracted using a program-verify approach from the pulse-training data of Figure~\ref{fig:2t1r_pulse}. (a)~Distributions for the HD SET (blue squares) and LD SET (brown circles) protocols at target levels of 30\,\textmu A--80\,\textmu A, with color indicating increasing target current. (b)~Coefficient of variation $C_V$ as a function of mean read current.}
    \label{fig:2t1r_MLC}
\end{figure}

\begin{figure}[tbh]
    \centering
    \includegraphics[width=0.6\linewidth]{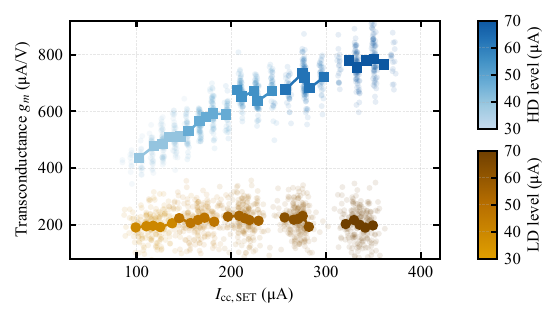}
    \caption{Effective transconductance $g_m$ of the HD (blue) and LD (brown) transistors as a function of SET compliance current $I_\mathrm{cc,SET}$, estimated as the incremental compliance current per unit gate voltage increment (equation~\ref{eq:gm}).}
    \label{fig:2t1r_algo}
\end{figure}

These results establish that the LD--HD configuration provides a favorable combination of fine programming granularity, reliable compliance current control, and low cycle-to-cycle variability. To further evaluate whether these intermediate states can be reliably targeted and distinguished under a practical multilevel programming scheme, six target read-current levels are defined, corresponding to read currents of 30\,\textmu A, 40\,\textmu A, 50\,\textmu A, 60\,\textmu A, 70\,\textmu A, and 80\,\textmu A. For each target level, all traces from Figure~\ref{fig:2t1r_pulse} are scanned to identify the first SET pulse at which the read current reaches or exceeds the target value, emulating a program-verify scheme. This leads to read current distributions for all six target levels, each containing only the read current recorded after the SET pulse at which the read current first reaches or exceeds the target value. Figure~\ref{fig:2t1r_MLC}~a) shows the resulting distributions for both the HD SET (blue squares) and LD SET (brown circles) protocols, with the color gradient from light to dark corresponding to increasing WL voltage. As a consequence of the program-verify methodology, the lower bound of each target level is strictly defined by the target current, whereas the upper bound is determined by the overshoot after the first successful SET pulse. However, a clear difference emerges in the upper bounds: the LD SET protocol produces narrower distributions, reducing the probability of overlap with the adjacent higher level. In particular, for the three lowest target levels of 30\,\textmu A, 40\,\textmu A, and 50\,\textmu A, the HD SET distributions of neighboring levels exhibit visible overlap, which would introduce ambiguity in state readout. In contrast, the corresponding LD SET distributions remain well separated between adjacent levels, with no measurable overlap. Figure~\ref{fig:2t1r_MLC}~b) presents the coefficient of variation $C_V$ at each target level, confirming that the LD SET protocol achieves lower variability than the HD SET protocol across all six defined levels. The difference in variability between the two protocols can be attributed to the difference in transconductance between the HD and LD transistors, which is analyzed in detail in Figure~\ref{fig:2t1r_algo}. The HD transistor exhibits a higher transconductance, such that an equivalent WL voltage increment produces a larger change in saturation current and consequently a larger step in read current. This increases the probability of overshooting the target level within a single SET pulse, broadening the upper tail of the distribution and increasing inter-level overlap. The LD transistor, by contrast, converts the same WL voltage increment into a smaller current step, enabling more precise targeting of each level and resulting in tighter distributions and lower $C_V$ observed across all target levels. Since the read current is determined by $I_\mathrm{cc}$ (see Figure~\ref{fig:2t1r_pulse}\,~c)), the wider programming window of the LD SET protocol (1.7\,V to 3.3\,V, 65 steps) further contributes to finer state resolution, making the LD--HD configuration of the 2T1R cell better suited for multilevel analog programming than configurations relying on a single transistor geometry for both programming and RESET.
To provide a transistor-level explanation for the superior programming granularity of the LD SET protocol, the effective 
transconductance of both transistors is analyzed. The transconductance is estimated as the incremental SET compliance current per unit WL voltage increment:
\begin{equation}
    g_m = \frac{I_\mathrm{cc,SET}^{(n+1)} - 
    I_\mathrm{cc,SET}^{(n)}}{\Delta V_{\mathrm{WL}}^{\mathrm{SET}}}
    \label{eq:gm}
\end{equation}
where $I_\mathrm{cc,SET}^{(n)}$ and $I_\mathrm{cc,SET}^{(n+1)}$ denote the compliance currents at consecutive gate voltage steps, and $\Delta V_{\mathrm{WL}}^{\mathrm{SET}} = 25\,\mathrm{mV}$ is the WL voltage increment. Figure~\ref{fig:2t1r_algo} presents the resulting $g_m$ as a function of $I_\mathrm{cc,SET}$ for both transistors. Blue squares and brown circles denote the HD and LD transistors, respectively. The HD transistor exhibits a significantly higher transconductance than the LD transistor at every target level, directly confirming that an equivalent gate voltage increment of $\Delta V_\mathrm{WL,SET} = 25\,\mathrm{mV}$ produces a larger change in $I_\mathrm{cc,SET}$ for the HD transistor. This translates into a larger read current step per pulse and broader inter-level distributions, consistent with the overlap observed in Figure~\ref{fig:2t1r_pulse}~a). The transconductance of the LD transistor remains approximately constant at around $200$\,\textmu A/V across all target levels, whereas the transconductance of the HD transistor increases with $I_\mathrm{cc,SET}$. At elevated gate voltages, the larger and increasing transconductance of the HD transistor makes precise resistance state targeting increasingly difficult, as each pulse increment produces a progressively larger current step. This suggests a natural division of roles between the two transistors: the LD transistor is better suited for fine-grained multilevel programming, while the HD transistor is more appropriate for coarse programming at higher current levels. From this, we propose exploiting this complementary behavior through a combined coarse--fine programming strategy.
\subsection{Content Addressable Memory (CAM)}
\label{sec:res_cam}

The CAM results presented in this section are based exclusively on circuit simulations. We evaluate the resulting dynamic range and impact of supply-voltage variations on the top- and bottom-connected memristive device-transistor configurations introduced in Section~\ref{sec:cam_intro}. To this end, the nominal supply voltage of $V_\mathrm{DD,nom}=0.60~\mathrm{V}$ was varied by $\pm 10\%$ while monitoring the resulting shift in $V_{\mathrm{DL,bound}}$, which defines the programmable decision boundary of the \ac{aCAM}. Since stable decision boundaries are essential for reliable associative search, the sensitivity of $V_{\mathrm{DL,bound}}$ to supply-voltage variations serves as a key indicator of \ac{aCAM} robustness.
\begin{figure}[htb!]
    \centering
    \includegraphics[width=0.85\linewidth]{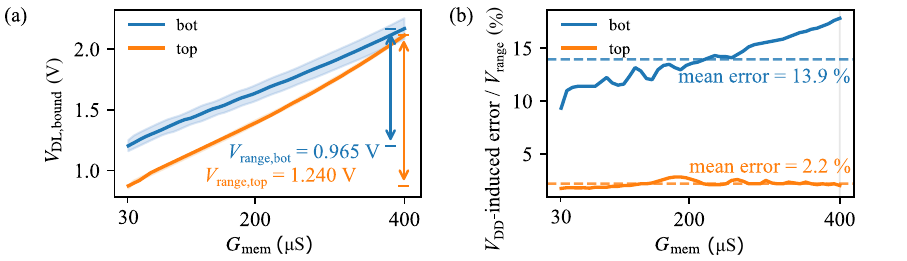}
    \caption{Simulated \ac{aCAM} dynamic range and impact of supply-voltage variation on the top- and bottom-connected RRAM comparator configurations. 
    (a)~$V_{\mathrm{DL,bound}}$ as a function of $G_{\mathrm{mem}}$ for the nominal supply voltage ($V_{\mathrm{DD,nom}} = 0.60\,\mathrm{V}$) and a $\pm 10\%$ supply-voltage variation. (b)~Corresponding supply-induced error normalized to the obtainable dynamic range.}
    \label{fig:cam_results}
\end{figure}
Figure~\ref{fig:cam_results} shows that the top-connected configuration achieves a larger obtainable dynamic range ($1.240~\mathrm{V}$ versus $0.965~\mathrm{V}$) while exhibiting significantly lower sensitivity to supply variations. The mean normalized $V_\mathrm{DD}$-induced error is reduced from $13.9\%$ for the bottom-connected configuration to $2.2\%$ for the top-connected configuration.

These results indicate that the top-connected architecture provides both a larger sensing margin $V_{\mathrm{range}}$ and improved robustness against supply-voltage fluctuations. Experimental chip measurements, currently underway, will determine whether these advantages outweigh potential trade-offs in memristive device programmability.

\subsection{Compute-in-Memory MAC Operation}

The multiply and accumulate (MAC) operation constitutes the fundamental computational primitive in neural-network accelerators. RRAM enables MAC operations by exploiting Kirchhoff's Current Law (KCL) and Kirchhoff's Voltage Law (KVL), allowing computations to be performed directly within the memory array and thereby minimizing costly data movement between memory and processing units. In this section, we demonstrate the hardware implementation of MAC operations in the fabricated $15 \times 15$ 2T1R CIM1 RRAM crossbar array and investigate the impact of different programming and read transistor configurations on MAC accuracy.

MAC operations were implemented using an FPGA-based memory characterization platform (Arc TWO, Arc Instruments). The architecture of the fabricated 2T1R CIM1 array is illustrated in Figure~\ref{fig:CIMdesign}~b). Devices within each column share a common bit line (BL), while devices within each row share a common source line (SL) and word line corresponding to the HD (WL1) and LD (WL2) access transistors. Programming and read operations are performed through appropriate biasing of the BLs, SLs, and WLs. During SET programming, a voltage $V_{\mathrm{SET}}$ is applied to the selected BL while the corresponding SL is grounded. Unselected BLs and SLs are held at ground potential to suppress undesired programming. The WL voltage of either the HD or LD transistor is raised to $V_{\mathrm{WL}}^{\mathrm{SET}}$, depending on the selected programming configuration, thereby defining the current compliance during filament formation. RESET operations are performed by reversing the polarity across the selected device and activating only the HD transistor, which provides sufficient drive current to ensure reliable RESET switching. Read operations follow a similar biasing scheme as SET programming but employ a lower read voltage $V_{\mathrm{READ}}$. 
Note that in this section, the X–Y notation refers to the transistor used for SET programming (X) and MAC readout (Y), respectively, distinct from the SET–RESET notation used in Section 6.1. To investigate the influence of access-transistor selection on MAC performance, four programming/MAC column READ configurations were evaluated: (i) HD--HD, where both SET programming and MAC readout employ the HD transistor; (ii) HD--LD, where SET programming uses HD and MAC readout uses LD; (iii) LD--HD, where SET programming uses LD and MAC readout uses HD; and (iv) LD--LD, where both operations use the LD transistor. RESET operations were always performed using the HD transistor to guarantee reliable RESET operation due to high drive-current capability of the HD transistor. For the MAC experiment, five distinct weight matrices were programmed: an all-ones matrix, an all-zeros matrix, two complementary checkerboard patterns, and one randomly generated pattern to cover all the corner cases. During the weight matrix programming, devices representing the LRS states were programmed to a target resistance of $3~\mathrm{k\Omega}$ using an incremental $V_{\mathrm{SET}}$ and $V_{\mathrm{WL}}$ program-verify algorithm with a tolerance window of $\pm1\%$. In cases of overshoot, corrective RESET pulses were applied to restore the target resistance. The $3~\mathrm{k\Omega}$ LRS target was selected to ensure that it could be reliably programmed using both the HD and LD transistors, given the limited drive current capability of the LD transistor. The HRS devices were programmed using an incremental RESET-and-verify procedure until a resistance exceeding $100~\mathrm{k\Omega}$ was achieved. All programming pulses had a fixed duration of \SI{1}{\micro\second}. During verification, device resistances were measured using the HD transistor with a gate voltage of 3.3~V and a read voltage of 0.2~V. 

MAC operations were subsequently performed using 1,000 randomly generated binary input vectors for each programmed weight matrix. The input vector was applied by selectively activating WLs according to the binary input pattern, while the programmed conductance states represented the weight matrix. The MAC operation is then performed by applying $V_{\mathrm{BL}}=V_{\mathrm{READ}}$, resulting in an accumulated current along the column which is then measured and taken as the measured MAC output. To evaluate MAC accuracy, the measured column current is compared with the corresponding ideal digital MAC output. For binary weight and input vectors, $\mathbf{w}=[w_1,w_2,\ldots,w_N]$ and $\mathbf{x}=[x_1,x_2,\ldots,x_N]$, respectively, where $w_i,x_i \in \{0,1\}$, the ideal digital MAC output is given by

\begin{equation}
\mathrm{MAC} = \sum_{i=1}^{N} w_i x_i ,
\label{eq:mac}
\end{equation}

which corresponds to the binary dot product between the input and weight vectors.
\begin{figure}[tbh]
    \centering
    \includegraphics[width=0.9\linewidth]{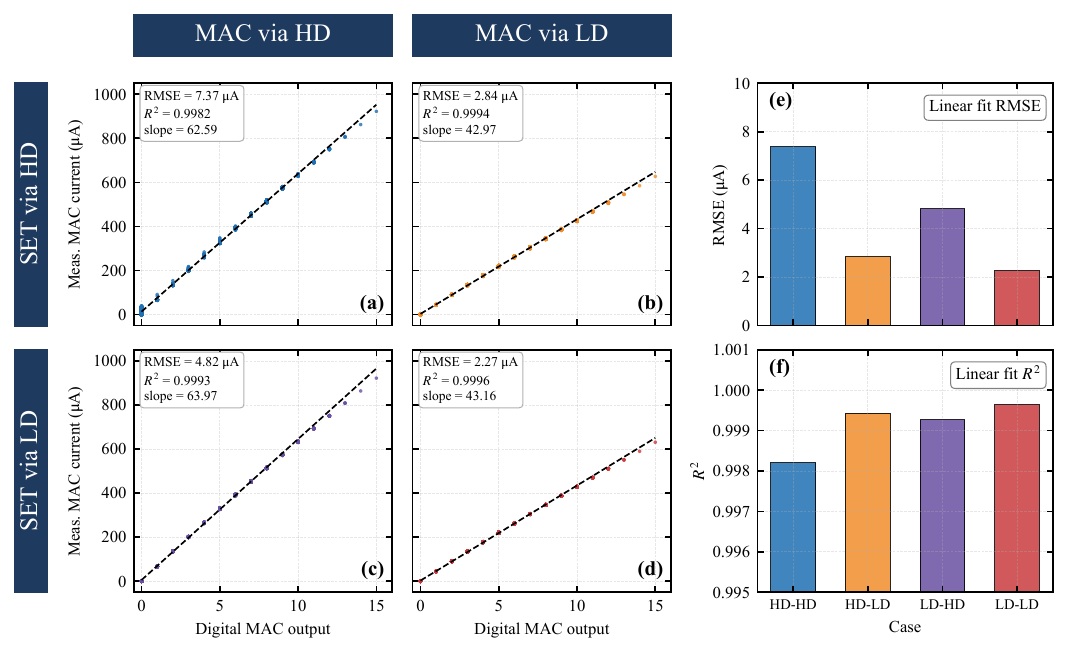}
    \caption{Compute-in-memory MAC operation on the 2T1R RRAM crossbar array. (a--d)~Measured MAC column current versus digital MAC output for 5{,}000 MAC operations for various binary input vectors and weight matrix combinations, shown for all four cases: HD--HD, HD--LD, LD--HD, and LD--LD respectively. The dashed line represents the linear fit.  (e)~Linear fit RMSE of the MAC output current for the four transistor combinations. (f)~Linear fit $R^2$ of the MAC output current for the four transistor combinations. The LD--LD case yields the most linear MAC response with the lowest RMSE error of 2.27 \si{\micro\ampere}.}
\label{fig:mac}
\end{figure}

To quantify MAC accuracy, the measured MAC current was plotted as a function of the ideal digital MAC output and fitted using a linear model. The quality of the MAC operation was evaluated using the mean absolute error (MAE), root-mean-square error (RMSE), and coefficient of determination ($R^2$) extracted from the linear fit. The MAE and RMSE quantify the deviation of the measured MAC current from the fitted linear response, while the $R^2$ value provides a measure of linearity between the measured MAC current and the corresponding digital MAC output. In addition, the fitted slope represents the average current contribution per active bit and serves as an indicator of the effective MAC gain. Figures~\ref{fig:mac}~a)--d) compare the measured MAC current with the corresponding ideal digital MAC output for the four programming/read configurations for 5,000 MAC operations. The extracted fitting parameters and error metrics, summarized in Table~\ref{tab:mac_accuracy}, enable a quantitative comparison of MAC accuracy and linearity. Figures~\ref{fig:mac}~e) and~f) further summarize the RMSE and $R^2$ values obtained for the four transistor combinations. Among the investigated configurations, LD--LD achieves the lowest RMSE of $2.27\,\mu\mathrm{A}$ and the highest $R^2$ value of 0.9996, indicating the best MAC accuracy and linearity. In contrast, the HD--HD configuration exhibits the highest RMSE of 7.37 \si{\micro\ampere} and the lowest $R^2$ value of 0.9982. The superior performance of the LD--LD configuration is attributed to its improved programming accuracy, as discussed in the previous section, together with reduced read-current variability resulting from the comparable on-resistance of the LD transistor and the programmed LRS of the RRAM device. Although this configuration yields a lower MAC gain, corresponding to a slope of 43.16 \si{\micro\ampere}/bit, the gain reduction can be readily compensated through calibration in the readout circuitry. These results demonstrate that the LD transistor is beneficial not only for achieving precise conductance programming but also for enabling highly accurate compute-in-memory MAC operations.

\begin{table}[t]
\centering
\caption{Comparison of MAC accuracy metrics for different SET/MAC READ transistor configurations.}
\label{tab:mac_accuracy}
\begin{tabular}{lcccc}
\toprule
Case & MAE (\si{\micro\ampere}) & RMSE (\si{\micro\ampere}) & $R^2$ & Slope (\si{\micro\ampere}/bit) \\
\midrule
HD--HD & 5.12 & 7.37 & 0.9982 & 62.59 \\
HD--LD & 2.37 & 2.84 & 0.9994 & 42.97 \\
LD--HD & 4.04 & 4.82 & 0.9993 & 63.97 \\
LD--LD & 1.89 & 2.27 & 0.9996 & 43.16 \\
\bottomrule
\end{tabular}
\end{table}

\section{Conclusion}
This work presented \textit{OTTER}, a 28\,nm CMOS-based research vehicle co-integrated with TaO$_\textrm{x}$-based RRAM, designed to systematically investigate the 2T1R architecture at both the device and array level. The chip integrates dedicated 2T1R test structures as well as content-addressable memory (CAM) and compute-in-memory (CIM) arrays, enabling a comprehensive evaluation of switching behavior and application-level functionality within a unified platform.\\
Through systematic experimental characterization of three transistor--pairing configurations, the 2T1R architecture is shown to provide independent control over SET and RESET current paths. The HD transistor provides reliable RESET across all bias condition, while the LD transistor enables finer programming granularity during SET. Design guidelines of transistor sizing are derived from the combination of DC sweep measurements and compact model simulations, establishing the minimum RESET transistor $W/L$ ratio required for complete RESET as a function of the SET current compliance.Under pulse-based multilevel programming, the lower transconductance of the LD transistor translates each gate voltage increment into a smaller compliance current step, reducing the probability of overshooting the target conductance state. As a result, the LD SET protocol produces tighter distributions at each target level with reduced inter-level overlap compared with the HD SET protocol, despite comparable intrinsic cycle to cycle variability.
Simulation-based analysis of an aCAM design further evaluates trade-offs between top- and bottom-connected RRAM comparator configurations on the same platform.
Exemplary MAC operations via the CIM1 pseudo-crossbar array demonstrate the versatility of the 2T1R architecture for in-memory computing. Among the four evaluated programming/readout transistor combinations, using the LD transistor for both SET programming and MAC readout achieves the lowest RMSE and highest linearity. In addition, a compact model calibrated to measured device characteristics was developed and used to reproduce the observed behavior, providing a valuable tool for circuit- and system-level simulations. The \textit{OTTER} chip therefore serves as both a validation platform for RRAM-CMOS co-integration in advanced technology nodes and a foundation for future neuromorphic systems, paving the way toward larger-scale implementations.

\funding
{This work was funded by the Federal Ministry of Research, Technology and Space (BMFTR, Germany) in the project NEUROTEC-II (project numbers: 16ME0398K and 16ME0399)}

\roles{Y.C.: Conceptualization, Investigation, Validation, Writing -- original draft. D.S.: Methodology, Investigation, Writing -- review \& editing. X.Z.: Methodology, Investigation, Writing -- review \& editing. A.B.: Investigation, Validation, Writing -- review \& editing. P.-P.M.: Conceptualization, Methodology, Investigation, Writing -- review \& editing. O.A.: Conceptualization. A.A.: Methodology, Writing -- review \& editing. K.W.: Methodology. G.P.: Conceptualization. S.J.: Conceptualization. C.R.: Validation. S.K.: Methodology. V.R.: Conceptualization. D.W.: Conceptualization, Methodology. M.S., S.M., A.Z., C.G., S.W., S.H.-E., J.P.S., and S.v.W.: Conceptualization, Supervision, Funding acquisition, Writing -- review \& editing. R.D.: Supervision, Funding acquisition, Project administration.}

\data{The data that support the findings of this study are available from the corresponding authors upon reasonable request.}

\suppdata{Supplementary data associated with this article include additional figures, tables, and methodological details and are available online with the published article.}

\printbibliography
\end{document}